\documentclass[showpacs,pra,superscriptaddress,twocolumn]{revtex4-1}
\usepackage{amsmath}
\usepackage{graphicx}
\usepackage{epsfig}
\usepackage{amsfonts}
\usepackage{color}
\usepackage{epstopdf}
\usepackage{hyperref}
\usepackage[T1]{fontenc}
\usepackage[latin9]{inputenc}
\usepackage{amsmath}
\usepackage{amssymb}
\usepackage{graphicx}
\usepackage{esint}
 \usepackage{footnote}

\begin{document}

\title{ Steady-State Coherences under Partial Collective non-Markovian Decoherence}

\author{S.~L.~Wu}
\thanks{These authors contributed to the work equally and should be regarded as co-first authors.}
\affiliation{School of Physics and Materials Engineering, Dalian Minzu University, Dalian 116600, China}

\author{W.~Ma}
\thanks{These authors contributed to the work equally and should be regarded as co-first authors.}
\affiliation{School of Physics and Materials Engineering, Dalian Minzu University, Dalian 116600, China}

\author{Zhao-Ming Wang}
\affiliation{College of Physics and Optoelectronic Engineering, Ocean University of China, Qingdao 266100, China}

\author{P.~Brumer}
\affiliation{Chemical Physics Theory Group, Department of Chemistry, and Center for Quantum Information and Quantum Control, University of Toronto, Toronto, Ontario, Canada M5S 3H6}

\author{Lian-Ao~Wu}
\email{lianao.wu@ehu.es}
\affiliation{Department of Physics, University of the Basque Country UPV/EHU, 48080 Bilbao, Spain}
\affiliation{IKERBASQUE, Basque Foundation for Science, Bilbao, 48011, Spain}
\affiliation{EHU Quantum Center, University of the Basque Country UPV/EHU, Leioa, Biscay 48940, Spain}

\date{\today}

\begin{abstract}
Steady-state coherence in open quantum systems is crucial for quantum technologies, yet its behavior is not fully understood due to the interplay between collective and individual decoherence. While collective decoherence is thought to induce steady-state coherence, experiments often fail to observe this because of individual decoherence. We study a system of two harmonic oscillators coupled to both individual and collective environments, introducing a tuneable parameter to adjust the decoherence proportions. By analytically solving the exact dynamical equations, we find that steady-state coherence depends on the initial state under collective decoherence, but not under partial decoherence. Interestingly, non-Markovianity induces rich and complex steady-state coherence behaviors. 
Our results offer new insights into the role of non-Markovian decoherence in quantum systems and serve as a benchmark for evaluating approximate methods in modeling quantum processes.
\end{abstract}

\pacs{03.67.-a, 03.65.Yz, 05.70.Ln, 05.40.Ca}

\maketitle

\section{Introduction}

The study of steady-state coherence in open quantum systems is
of fundamental importance in quantum physics and has practical
implications for quantum technologies such as quantum computation
\cite{Verstraete2009,Yosifov2024},  quantum metrology
\cite{Wang2018,Naghiloo2019},
and quantum thermodynamics\cite{Goold2016,Lu2024}. Coherence,
as a signature of quantum superposition, is essential for these
applications, but its feature in the presence of environmental
couplings is a major challenge \cite{Ask2022,Ablimit2024,Singh2021}.
Understanding how coherence behaves in the long-time limit, especially
in systems coupled with multiple or various environments simultaneously
\cite{Becker2022,Tucker2020}, is critical for the development of
robust and efficient quantum systems. This makes the investigation
of steady-state coherence not only theoretically significant but also
practically necessary.

Theoretically, under complete collective decoherence \cite{Liu2024,Venkatesh2018},
the common environment can  induce steady-state coherence in
an open quantum system\cite{Guarnieri2018,Gribben2020,Zhou2022}.
This intriguing prediction suggests that even in the presence of dissipation,
coherence can be preserved or even generated in certain    conditions
\cite{Zanardi2014,Dodin2021,Dodin2024}. This forms the basis for the concept of the decoherence-free
subspace \cite{Venkatesh2018, Wu2002}. However,
experimental observations often fail to confirm such predictions \cite{Pushin2011},
with feeble steady-state coherence being detected. The primary reason for this
discrepancy lies in the unavoidable presence of individual decoherence
\cite{Mewes2005,Ruimy2021}, which competes with and suppresses the
effects of collective decoherence. Given the diversity of physical systems
and environments, it is essential to study models that incorporate both
collective and individual decoherence to better understand their combined
effects and the conditions under which steady-state coherence might survive
\cite{Eastham2016}.


{
Exactly solvable models are rare but highly valuable, as they offer precise and detailed insights into new concepts in physics. They can serve as benchmarks, helping us determine when approximate equations remain valid. For example, bilinear models of identical particles lead to integro-differential equations that can be solved exactly for general spectra (such as Ohmic spectra) \cite{PhysRevD.45.2843,PhysRevE.55.153,
PhysRevLett.109.170402,arXiv:2211.15722}. Analytical solutions of such models are particularly important because they give clear physical insight into the nature of open-system dynamics. Solutions have been obtained for a single harmonic oscillator coupled to a bosonic reservoirs with specific spectral densities~\cite{PhysRevA.32.2462, PhysRevA.111.062206}. However, analytical solutions for open two-oscillator systems, which are essential for studying exact coherence and entanglement, are still missing.}

In this work we focus on the steady-state coherence for a pair of non-interacting harmonic oscillators that
are simultaneously coupled to {individual} and {collective} bosonic reservoirs.
By introducing a tunable parameter that continuously interpolates between the
two decoherence channels~\cite{Wang2023}, we derive exact Heisenberg-picture
equations of motion, solve them analytically for a Lorentzian (Drude-Lorentz)
spectral density, and identify the conditions under which steady-state coherence
survives or even increases~\cite{Braun2011}.
Our results reveal that the balance between collective and individual
decoherence critically determines the steady-state coherence of the system.
Using the main values of suitable observables to quantify the coherence
of the harmonic oscillators\cite{Chou2008}, we find that the properties of steady-state
coherence under perfect collective decoherence differ fundamentally from
those under partial collective decoherence. Remarkably, under collective
decoherence, the steady-state coherence depends on the system's initial
state, while partial collective decoherence suppresses this dependence.
Additionally, contrary to previous results, Even for the partial collective decoherence model,
by appropriately selecting system parameters, we can always ensure the observation of
steady-state quantum coherence, especially in the non-Markovian regime
\cite{Ask2022,Ablimit2024}. These findings provide a deeper
understanding of how environmental interactions influence quantum
coherence and highlight the necessity of considering both types of
decoherence in realistic models.

This paper is organized as follows: In Sec. \ref{ED}, we present the exact dynamical
equations of the harmonic oscillator system and the analytical solutions in the
Heisenberg picture { when their reservoir spectral densities are Lorentzian}. In Section \ref{NR}, we present numerical results illustrating the
steady-state coherence under varying conditions. Conclusions are presented in Sec.
\ref{Conc}, { Conclusions are presented in Sec. IV, where we illustrate approximate methods such as reaction coordinate (RC) mapping ~\cite{Strasberg2016,Hartmann2020,Weiss2021} and qualitatively discuss our analytical solutions and their roles in evaluating approximations, for example, as benchmarks.}

\section{Exact Dynamics}\label{ED}

\subsection{Exact Dynamical Equation}

We consider a system consisting of two harmonic oscillators, each coupled to its own bosonic reservoir at finite temperatures $T_i$ ($i=1,2$) and to a shared common bosonic heat reservoir at temperature $T_0$. The total Hamiltonian is given by
\[
\hat{H}_{\text{tot}} = \hat{H}_{\text{S}} + \hat{H}_{\text{B}} + \hat{H}_{\text{SB}},
\]
where the bath Hamiltonian reads
\[
\hat{H}_{\text{B}} = \sum_{i=0,1,2} \sum_{\alpha} \omega_{i,\alpha} \hat{b}_{i,\alpha}^\dagger \hat{b}_{i,\alpha},
\]
with $\hat{b}_{i,\alpha}$ the annihilation operator for the $i$th reservoir mode of frequency $\omega_{i,\alpha}$. The system Hamiltonian is
\[
\hat{H}_{\text{S}} = \sum_{i=1,2} \omega_i \hat{c}_i^\dagger \hat{c}_i,
\]
where $\hat{c}_i$ annihilates a quantum in oscillator $i$. The interaction Hamiltonian is
\[
\hat{H}_{\text{SB}} = \sum_{i=0,1,2} \sum_{\alpha} g_{i,\alpha} \hat{L}_i \otimes \hat{b}_{i,\alpha}^\dagger + \text{h.c.},
\]
where the system coupling operators are defined as $\hat{L}_0 = \hat{c}_1 + \hat{c}_2$ for the common reservoir and $\hat{L}_i = \hat{c}_i$ for the individual reservoirs ($i=1,2$), with $g_{i,\alpha}$ the coupling strengths.

To describe the dynamics, we start from the Heisenberg equations of motion for the system operators, which read
\begin{eqnarray}
\dot{\hat{c}}_1 &=& -i \omega_1 \hat{c}_1 - i \sum_{\alpha} g_{0,\alpha}^* \hat{b}_{0,\alpha} - i \sum_{\beta} g_{1,\beta}^* \hat{b}_{1,\beta}, \label{eq:c0-1} \\
\dot{\hat{c}}_2 &=& -i \omega_2 \hat{c}_2 - i \sum_{\alpha} g_{0,\alpha}^* \hat{b}_{0,\alpha} - i \sum_{\beta} g_{2,\beta}^* \hat{b}_{2,\beta}. \label{eq:c1-1}
\end{eqnarray}
Similarly, the reservoir mode operators evolve according to
\begin{eqnarray}
\dot{\hat{b}}_{i,\alpha} &=& -i \omega_{i,\alpha} \hat{b}_{i,\alpha} - i g_{i,\alpha} \hat{L}_i. \label{eq:bi-1}
\end{eqnarray}
We formally integrate Eq.~(\ref{eq:bi-1}) to express the bath operators as a function of their initial values and the system operators:
\begin{eqnarray*}
\hat{b}_{i,\alpha}(t) &=& \hat{b}_{i,\alpha}(0) e^{-i \omega_{i,\alpha} t} \\
&& - i g_{i,\alpha} \int_0^t ds\, \hat{L}_i(s) e^{-i \omega_{i,\alpha} (t-s)}.
\end{eqnarray*}
Substituting this solution back into Eqs.~(\ref{eq:c0-1}) and (\ref{eq:c1-1}) yields integro-differential equations for the system operators, explicitly incorporating the memory effects of the reservoirs:
\begin{eqnarray*}
\dot{\hat{c}}_j(t) &=& -i \omega_j \hat{c}_j - i \sum_{\alpha} g_{0,\alpha}^* \hat{b}_{0,\alpha}(0) e^{-i \omega_{0,\alpha} t} \\
&& - \sum_{\alpha} |g_{0,\alpha}|^2 \int_0^t ds\, \hat{L}_0(s) e^{-i \omega_{0,\alpha}(t-s)} \\
&& - \sum_{\beta} |g_{j,\beta}|^2 \int_0^t ds\, \hat{c}_j(s) e^{-i \omega_{j,\beta}(t-s)} \\
&& - i \sum_{\beta} g_{j,\beta}^* \hat{b}_{j,\beta}(0) e^{-i \omega_{j,\beta} t}.
\end{eqnarray*}

We now seek a solution by expressing $\hat{c}_j(t)$ as a linear combination of the initial system and bath operators, which reflects how information about the initial state propagates under the open system dynamics:
\begin{eqnarray}
\hat{c}_j(t) &=& \sum_i C_i^j(t) \hat{c}_i(0) + \sum_{k,\alpha} B_{k,\alpha}^j(t) \hat{b}_{k,\alpha}(0), \label{eq:expsa}
\end{eqnarray}
where $C_i^j(t)$ and $B_{k,\alpha}^j(t)$ are time-dependent coefficients encoding the influence of initial conditions.
The exact dynamical equations governing these coefficients follow directly from the Heisenberg equations:
\begin{eqnarray}
\dot{C}_i^j(t) &=& -i \omega_j C_i^j(t) - \int_0^t d\tau\, f_j(t-\tau) C_i^j(\tau) \nonumber \\
&& - \int_0^t d\tau\, f_0(t-\tau) \sum_p C_i^p(\tau), \label{eq:Cij} \\
\dot{B}_{k,\alpha}^j(t) &=& -i \omega_j B_{k,\alpha}^j(t) - \int_0^t d\tau\, f_j(t-\tau) B_{k,\alpha}^j(\tau) \nonumber \\
&& - \int_0^t d\tau\, f_0(t-\tau) \sum_{p=1,2} B_{k,\alpha}^p(\tau) \nonumber \\
&& - i g_{k,\alpha}^* e^{-i \omega_{k,\alpha} t} \delta_{j k} - i g_{0,\alpha}^* e^{-i \omega_{0,\alpha} t} \delta_{k0}, \label{eq:Bij}
\end{eqnarray}
with memory kernels defined as $f_k(t-s) = \sum_\alpha |g_{k,\alpha}|^2 e^{-i \omega_{k,\alpha}(t-s)}$ for $k=0,1,2$.

Once these coefficients are determined, the time evolution of any observable can be computed straightforwardly, as all operators are expressed explicitly in terms of their initial conditions. This formulation highlights how the interplay of individual and collective decoherence channels is embedded in the system dynamics through the structure of these coefficients.

\subsection{Analytical Solutions}

Under certain symmetry conditions, the dynamical equations of the open oscillator system admit an analytical solution. Specifically, we assume identical oscillator frequencies and reservoir modes, i.e., $\omega_1 = \omega_2$, $\omega_{1,\alpha} = \omega_{2,\alpha}$, and symmetric coupling strengths $g_{1,\alpha} = g_{2,\alpha}$. We parametrize the system-bath couplings as $g_{0,\alpha} = g_\alpha \cos^2\theta$ for the common bath and $g_{1,\alpha} = g_\alpha \sin^2\theta$ for the individual baths, where $g_\alpha$ is an effective overall coupling strength and $\theta$ is a tunable parameter that controls the relative contributions of individual and collective decoherence.
This parametrization allows us to interpolate continuously between two limiting cases: for $\theta = \pi/2$, the system is coupled completely to individual reservoirs (completely individual decoherence), whereas for $\theta = 0$, the system interacts completely with a common environment (completely collective decoherence)~\cite{Wang2023}.

To proceed towards an exact solution, we group the coefficients \(C_j^i\) and \(B_j^i\) and introduce suitable transformations to simplify the coupled integro-differential equations into a set of decoupled second-order differential equations for auxiliary variables.
As an illustrative example, we focus on the group of coefficients \(C_j^i\), whose evolution equations take the following form:
\begin{eqnarray*}
\dot{C}_{j}^{1}(t) & = & -i\,\omega_{1}C_{j}^{1}(t)-\sin^{4}\theta\int_{0}^{t}d\tau\,f(t-\tau)C_{j}^{1}(\tau)\\
&& -\cos^{4}\theta\int_{0}^{t}d\tau\,f(t-\tau)\sum_{p=1,2}C_{j}^{p}(\tau), \\
\dot{C}_{j}^{2}(t) & = & -i\,\omega_{1}C_{j}^{2}(t)-\sin^{4}\theta\int_{0}^{t}d\tau\,f(t-\tau)C_{j}^{2}(\tau)\\
&& -\cos^{4}\theta\int_{0}^{t}d\tau\,f(t-\tau)\sum_{p=1,2}C_{j}^{p}(\tau).
\end{eqnarray*}
These two equations describe the coupled memory-dependent dynamics of the two oscillators' coefficients, reflecting the interplay between collective and individual decoherence channels.
To decouple the above equations, we define symmetric and antisymmetric combinations of \(C_j^1\) and \(C_j^2\) as:
\begin{eqnarray*}
A_{j}^{1} & = & C_{j}^{1}+C_{j}^{2}, \\
A_{j}^{2} & = & C_{j}^{1}-C_{j}^{2}.
\end{eqnarray*}
Substituting these definitions leads to the following decoupled equations:
\begin{eqnarray}
\dot{A}_{j}^{1} & = & -i\,\omega_{1}A_{j}^{1} - \left(\sin^{4}\theta+2\cos^{4}\theta\right) \int_{0}^{t}d\tau\,f(t-\tau)A_{j}^{1}(\tau), \nonumber \\
\dot{A}_{j}^{2} & = & -i\,\omega_{1}A_{j}^{2} - \sin^{4}\theta \int_{0}^{t}d\tau\,f(t-\tau)A_{j}^{2}(\tau). \label{eq:dA1}
\end{eqnarray}
These transformed equations show that \(A_j^1\) and \(A_j^2\) evolve independently, with effective memory kernels that depend on \(\theta\).
Differentiating both equations once more yields second-order differential equations that are more convenient for analytical treatment:
\begin{eqnarray}
\ddot{A}_{j}^{1} & = & -i\,\omega_{1}\dot{A}_{j}^{1} - \left(\sin^{4}\theta+2\cos^{4}\theta\right) f(0) A_{j}^{1} \nonumber \\
&& - \left(\sin^{4}\theta+2\cos^{4}\theta\right) \int_{0}^{t}d\tau\,\dot{f}(t-\tau)A_{j}^{1}(\tau), \label{eq:dA2} \\
\ddot{A}_{j}^{2} & = & -i\,\omega_{1}\dot{A}_{j}^{2} - \sin^{4}\theta \bigg[ f(0)A_{j}^{2} - \int_{0}^{t}d\tau\,\dot{f}(t-\tau)A_{j}^{2}(\tau) \bigg]. \nonumber
\end{eqnarray}

To obtain an analytical solution, we choose a spectral density of the Lorentzian form:
\[
J(\omega) = \frac{\Gamma}{2\pi} \frac{\gamma^{2}}{(\omega - \omega_0)^2 + \gamma^2},
\]
where $\Gamma$ denotes the overall system-environment coupling strength, $\gamma$ is the spectral width (inverse correlation time), and $\omega_0$ is the central frequency of the spectrum.
This spectral density leads to an exponentially decaying correlation function:
\[
f(t-s) = \frac{\Gamma\,\gamma}{2} \exp\left[ -(\gamma + i\omega_0) |t-s| \right],
\]
which satisfies the differential relation
\[
\dot{f}(t-s) = -(\gamma + i\omega_0) f(t-s).
\]
Substituting this property into Eq.~(\ref{eq:dA2}), the integro-differential equations for $A_j^1$ and $A_j^2$ simplify to the following form:
\begin{eqnarray*}
\ddot{A}_j^1 &=& -i\omega_1 \dot{A}_j^1 - \left( \sin^4\theta + 2\cos^4\theta \right) f(0) A_j^1 \\
&& + \left( \sin^4\theta + 2\cos^4\theta \right) (\gamma + i\omega_0) \int_0^t d\tau\, f(t-\tau) A_j^1(\tau), \\
\ddot{A}_j^2 &=& -i\omega_1 \dot{A}_j^2 - \sin^4\theta f(0) A_j^2 \\
&& + \sin^4\theta (\gamma + i\omega_0) \int_0^t d\tau\, f(t-\tau) A_j^2(\tau).
\end{eqnarray*}
Next, we use Eq.~(\ref{eq:dA1}) to eliminate the remaining integral terms. Specifically, the memory integrals can be expressed as:
\begin{eqnarray*}
\int_0^t d\tau\, f(t-\tau) A_j^1(\tau) &=& -\frac{\dot{A}_j^1 + i\omega_1 A_j^1}{\sin^4\theta + 2\cos^4\theta}, \\
\int_0^t d\tau\, f(t-\tau) A_j^2(\tau) &=& -\frac{\dot{A}_j^2 + i\omega_1 A_j^2}{\sin^4\theta}.
\end{eqnarray*}
Inserting these expressions leads to two decoupled second-order homogeneous differential equations for \(A_j^1\) and \(A_j^2\):
\begin{eqnarray}
\ddot{A}_j^1 + y_1\,\dot{A}_j^1 + y_2\,A_j^1 &=& 0, \label{eq:2A1} \\
\ddot{A}_j^2 + y_1\,\dot{A}_j^2 + y_3\,A_j^2 &=& 0, \label{eq:2A2}
\end{eqnarray}
where the coefficients \(y_1, y_2, y_3\) encapsulate the effects of the reservoir spectral properties and decoherence structure:
\begin{eqnarray*}
y_1 &=& i(\omega_0 + \omega_1) + \gamma, \\
y_2 &=& i\omega_1 \gamma + \left[ \left( \sin^4\theta + 2\cos^4\theta \right) \frac{\Gamma\gamma}{2} - \omega_1 \omega_0 \right], \\
y_3 &=& i\omega_1 \gamma + \left[ \sin^4\theta \frac{\Gamma\gamma}{2} - \omega_1 \omega_0 \right].
\end{eqnarray*}
These second-order differential equations describe damped oscillator dynamics with effective damping and frequency renormalization determined by $\theta$. Importantly, both equations are analytically solvable, providing exact solutions for the decoupled variables \(A_j^1\) and \(A_j^2\), which fully characterize the system's evolution under Lorentzian spectral density.

{
While the dynamical equations up to Eq.~(8) are derived in a fully general form without specifying the spectral density, analytical progress beyond this point depends critically on the structure of the bath correlation function. Specifically, for a general spectral density \(J(\omega)\), the time derivative \(\dot{f}(t-s)\) cannot in general be expressed as a simple function of \(f(t-s)\) itself. As a result, the memory integrals in Eq.~(8) cannot be recast into local differential equations in \(A_j^i\), precluding a closed-form analytical solution. The Lorentzian case considered here is exceptional in that its exponential correlation function allows for this simplification.}

By following the same procedure used for \(A_j^i\), we can derive the differential equations governing the coefficients \(B_{j,\alpha}^i\). To simplify the resulting coupled equations, we introduce symmetric and antisymmetric combinations:
\begin{eqnarray*}
D_{j,\alpha}^{1} &=& B_{j,\alpha}^{1} + B_{j,\alpha}^{2}, \\
D_{j,\alpha}^{2} &=& B_{j,\alpha}^{1} - B_{j,\alpha}^{2}.
\end{eqnarray*}
These new variables satisfy the following second-order differential equations with inhomogeneous terms:
\begin{eqnarray}
\ddot{D}_{j,\alpha}^{1} + y_1\,\dot{D}_{j,\alpha}^{1} + y_2\,D_{j,\alpha}^{1} &=& y_j^1, \label{eq:Dj1} \\
\ddot{D}_{j,\alpha}^{2} + y_1\,\dot{D}_{j,\alpha}^{2} + y_3\,D_{j,\alpha}^{2} &=& y_j^2, \label{eq:Dj2}
\end{eqnarray}
where the explicit forms of the inhomogeneous terms are given by:
\begin{eqnarray*}
y_0^1 &=& -2\,\cos^2\theta\,g_\alpha^* e^{-i\omega_\alpha t} (\omega_\alpha - \omega_0 + i\gamma), \\
y_0^2 &=& 0, \\
y_1^1 &=& y_1^2 = y_2^1 = -y_2^2 = -(\omega_\alpha - \omega_0 + i\gamma) \sin^2\theta\,g_\alpha^* e^{-i\omega_\alpha t}.
\end{eqnarray*}
The detailed derivation of these equations is provided in Appendix~\ref{Dij}.

It is noteworthy that the second-order differential equations for \(A_j^i\) and \(D_{j,\alpha}^i\) share a common structure and can be written in the unified form:
\begin{eqnarray}
\ddot{x} + (a + i b)\,\dot{x} + (c + i d)\,x = h\,e^{-i\omega_\alpha t},\label{eq:GDE}
\end{eqnarray}
where \(x\) denotes a generic variable (\(A_j^i\) or \(D_{j,\alpha}^i\)) and \(h\) encodes the driving from the reservoir initial conditions.
The general solution consists of two parts, corresponding to the homogeneous and particular solutions:
\[
x(t) = x_h(t) + x_p(t).
\]
The homogeneous solution is determined by solving the characteristic equation:
\[
\lambda = \frac{-(a + i b) \pm \sqrt{(a + i b)^2 - 4(c + i d)}}{2},
\]
yielding roots \(\lambda_{1,2} = \alpha \pm i \beta\), where \(\alpha, \beta \in \mathbb{R}\) depend on the coefficients \(a, b, c, d\).
The general solution to the homogeneous equation then takes the form:
\[
x_h(t) = e^{\alpha t} \left[ C_1 \cos(\beta t) + C_2 \sin(\beta t) \right],
\]
where \(C_1\) and \(C_2\) are constants determined by the initial conditions.
The initial conditions for \(A_j^i\) and \(D_{j,\alpha}^i\) follow from the definitions of the expansion coefficients: initially, \(C_j^j(0) = 1\) and all other coefficients vanish.
For the inhomogeneous part, since the driving term is proportional to \(e^{-i\omega_\alpha t}\), a natural ansatz for the particular solution is:
\[
x_p(t) = A\,e^{-i\omega_\alpha t},
\]
where \(A\) is a complex constant to be determined. Substituting this ansatz into the differential equation yields:
\[
A = \frac{h}{c + b\,\omega_\alpha - \omega_\alpha^2 + i(d - a\,\omega_\alpha)},
\]
which provides an explicit solution for the particular solution.
This unified treatment highlights that both sets of coefficients \(A_j^i\) and \(D_{j,\alpha}^i\) can be solved analytically using the same mathematical structure, emphasizing the generality and robustness of the method employed.

The general solution to Eq.~(\ref{eq:GDE}) takes the form:
\begin{eqnarray}
x(t) &=& e^{\alpha t} \left[ C_1 \cos(\beta t) + C_2 \sin(\beta t) \right] \nonumber \\
&& + \frac{h\,e^{-i\omega_\alpha t}}{c + b\,\omega_\alpha - \omega_\alpha^2 + i(d - a\,\omega_\alpha)}, \label{eq:dx}
\end{eqnarray}
where the first term corresponds to the general solution of the homogeneous equation and the second term represents a particular solution driven by the reservoir initial conditions.
Once the constants \(C_1\) and \(C_2\) are determined from the initial conditions, this expression provides the complete analytical solution for the system dynamics.
However, our primary interest lies in the system's steady-state properties rather than its full time-dependent evolution. When individual decoherence is present (\(|\theta| \neq 0\)), it can be shown that \(\alpha < 0\), implying that the homogeneous solution decays exponentially with time.
Therefore, in the long-time limit, the first term in Eq.~(\ref{eq:dx}) vanishes and the steady-state behavior of the system is entirely governed by the particular solution \(x_p(t)\). This simplification allows us to extract steady-state observables directly from the particular solution, bypassing the need to track the full transient dynamics.

Since \(A_j^i\) satisfies a homogeneous second-order differential equation, all coefficients \(C_j^i\) decay to zero as the system relaxes to its steady state, reflecting the absence of external driving for this component.
In contrast, the coefficients \(D_{j,\alpha}^i\) obey inhomogeneous second-order differential equations [see Eqs.~(\ref{eq:Dj1}) and (\ref{eq:Dj2})], where the inhomogeneous terms originate from the initial conditions of the bath modes. As a result, \(D_{j,\alpha}^i\) generally approach nonzero steady-state values.
By performing straightforward calculations, these steady-state values can be obtained explicitly:

\begin{widetext}
\begin{eqnarray*}
B_{0,\alpha,\text{SS}}^{1}&=&B_{0,\alpha,\text{SS}}^{2}=\frac{-\mathrm{e}^{-i\,\omega_{\alpha}\,t}\,g_\alpha\,\cos^2\theta\,\left(\omega_{\alpha}-\omega_{0}+i\,\gamma\right)}{K_{c}},\\
B_{1,\alpha,\text{SS}}^{1}&=&\frac{-\mathrm{e}^{-i\,\omega_{\alpha}\,t}\,g_\alpha\,\sin\theta\,\left(\omega_{\alpha}-\omega_{0}+i\,\gamma\right)}{2}\left(\frac{1}{K_{s}}+\frac{1}{K_{c}}\right),\\
B_{1,\alpha,\text{SS}}^{2}&=&\frac{\mathrm{e}^{-i\,\omega_{\alpha}\,t}\,g_\alpha\,\sin\theta\,\left(\omega_{\alpha}-\omega_{0}+i\,\gamma\right)}{2}\left(\frac{1}{K_{s}}-\frac{1}{K_{c}}\right),\\
B_{2,\alpha,\text{SS}}^{1}&=&\frac{\mathrm{e}^{-i\,\omega_{\alpha}\,t}\,g_\alpha\,\sin\theta\,\left(\omega_{\alpha}-\omega_{0}+i\,\gamma\right)}{2}\left(\frac{1}{K_{s}}-\frac{1}{K_{c}}\right),\\\\
B_{2,\alpha,\text{SS}}^{2}&=&\frac{-\mathrm{e}^{-i\,\omega_{\alpha}\,t}\,g_\alpha\,\sin\theta\,\left(\omega_{\alpha}-\omega_{0}+i\,\gamma\right)}{2}\left(\frac{1}{K_{s}}+\frac{1}{K_{c}}\right).
\end{eqnarray*}
with
\begin{eqnarray*}
K_{c}&=&\frac{\Gamma\,\gamma}{2}\,\left(2\,\cos^4\theta+\sin^4\theta\right)-\omega_{\alpha}^{2}-\omega_{0}
\,\omega_{1}+\omega_{\alpha}\,\omega_{0}+\omega_{\alpha}\,\omega_{1}+i\,\gamma\,\left(\omega_{1}-
\omega_{\alpha}\right),\\
K_{s}&=&\frac{\Gamma\,\gamma}{2}\,\sin^{4}\theta-\omega_{\alpha}^{2}-
\omega_{0}\,\omega_{1}+\omega_{\alpha}\,\omega_{0}+\omega_{\alpha}\,\omega_{1}\,+i\,\gamma\,
\left(\omega_{1}-\omega_{\alpha}\right).
\end{eqnarray*}
\end{widetext}
For the general case, although a simple closed-form analytical solution is difficult to obtain, Eqs.~(\ref{eq:Cij}) and (\ref{eq:Bij}) can nevertheless be solved directly using the Laplace transform method. The detailed procedure is presented in Appendix~\ref{Lap}, ensuring that the full dynamical behavior of the system remains accessible beyond the analytically solvable cases discussed above.

\subsection{The Coherence Measurement}

In what follows, we evaluate the expectation values of the observables \(\langle J_x(t) \rangle\), \(\langle J_y(t) \rangle\), and \(\langle J_z(t) \rangle\), defined by
\[
\langle J_k(t) \rangle = \text{Tr} \{ J_k(t) \rho_{\text{tot}} \},
\]
where \(J_x(t)\) and \(J_y(t)\) correspond to the real and imaginary parts of the coherence \(\langle \hat{c}_1^\dagger \hat{c}_2 \rangle\) and thus serve as direct measures of coherence between the two oscillators~\cite{Chou2008}. The observable \(J_z(t)\) represents the population difference between the two oscillators.
We assume that initially the system and all reservoirs are uncorrelated, so that the total density matrix factorizes as
\[
\rho_{\text{tot}} = \rho_S(0) \otimes \rho_B^0(0) \otimes \rho_B^1(0) \otimes \rho_B^2(0),
\]
where each reservoir is prepared in a thermal equilibrium state at temperature \(T\):
\[
\rho_B^k(0) = \sum_n \frac{e^{-\beta \omega_{k,\alpha} n}}{Z_{k,\alpha}} |n_{k,\alpha}\rangle \langle n_{k,\alpha}|,
\]
with inverse temperature \(\beta = 1/T\), partition function \(Z_{k,\alpha} = 1/(1 - e^{-\beta \omega_{k,\alpha}})\), and Fock state \(|n_{k,\alpha}\rangle\). For simplicity, we assume all reservoirs are at the same temperature: \(T = T_0 = T_1 = T_2\).

Given this setup, the expectation values of the observables can be expressed as:{
\begin{eqnarray}
\langle J_x \rangle &=& \langle c_2^\dagger c_1 + c_1^\dagger c_2 \rangle \nonumber \\
&=& \sum_{i,j=1,2} \left( C_i^{2*} C_j^1 + C_i^{1*} C_j^2 \right) \langle \hat{c}_i^{0\dagger} \hat{c}_j^0 \rangle\nonumber \\
&& + \int d\omega\, J(\omega) \mathcal{B}_x(\omega) \frac{1}{e^{\beta \omega} - 1}, \label{eq:Jx}
\end{eqnarray}}
\begin{eqnarray*}
\langle J_y \rangle &=& i \langle c_1^\dagger c_2 - c_2^\dagger c_1 \rangle \\
&=& \sum_{i,j=1,2} i \left( C_i^{2*} C_j^1 - C_i^{1*} C_j^2 \right) \langle \hat{c}_i^{0\dagger} \hat{c}_j^0 \rangle \\
&& + i \int d\omega\, J(\omega) \mathcal{B}_y(\omega) \frac{1}{e^{\beta \omega} - 1},
\end{eqnarray*}
\begin{eqnarray*}
\langle J_z \rangle &=& \langle \hat{c}_2^\dagger \hat{c}_2 - \hat{c}_1^\dagger \hat{c}_1 \rangle \\
&=& \sum_{i,j=1,2} \left( C_j^{2*} C_i^2 - C_j^{1*} C_i^1 \right) \langle \hat{c}_j^{0\dagger} \hat{c}_i^0 \rangle \\
&& + \int d\omega\, J(\omega) \mathcal{B}_z(\omega) \frac{1}{e^{\beta \omega} - 1},
\end{eqnarray*}
where the functions \(\mathcal{B}_x\), \(\mathcal{B}_y\), and \(\mathcal{B}_z\) encode the reservoir contributions:
\begin{eqnarray*}
\mathcal{B}_x &=& \sum_i \left( \tilde{B}_{i,\alpha}^{1*} \tilde{B}_{i,\alpha}^2 + \tilde{B}_{i,\alpha}^{2*} \tilde{B}_{i,\alpha}^1 \right), \\
\mathcal{B}_y &=& \sum_i \left( \tilde{B}_{i,\alpha}^{2*} \tilde{B}_{i,\alpha}^1 - \tilde{B}_{i,\alpha}^{1*} \tilde{B}_{i,\alpha}^2 \right), \\
\mathcal{B}_z &=& \sum_i \left( |\tilde{B}_{i,\alpha}^2|^2 - |\tilde{B}_{i,\alpha}^1|^2 \right),
\end{eqnarray*}
with normalization \(\tilde{B}_{i,\alpha}^1(t) = B_{i,\alpha}^1(t) / g_{i,\alpha}\).
{After taking the continuum limit for the reservoir modes, the discrete mode index \(\alpha\) is replaced by the continuous frequency variable \(\omega\). In this context, \(\mathcal{B}_x(\omega)\) describes the contribution from bath modes at frequency \(\omega\). This replacement follows the standard procedure:
\[
\sum_\alpha |g_\alpha|^2 \rightarrow \int d\omega\, J(\omega),
\]
and does not involve a Fourier transformation. Consistently, in the continuum limit the bath average excited number becomes
\[
\langle \hat{b}_{i,\alpha}^{0\dagger} \hat{b}_{i,\alpha}^0 \rangle_B = \frac{1}{e^{\beta \omega} - 1}.
\]
}
This formulation allows for direct numerical evaluation of the relevant observables while clearly separating the contributions from initial system coherence and thermal reservoir fluctuations.

\section{The Steady State Coherences} \label{NR}

In this section, we investigate the steady-state characteristic of a double-resonator
system under different decoherence proportions by simulating the
exact dynamical equations [Eqs. (\ref{eq:Cij}) and (\ref{eq:Bij})].
From the analytical solutions, we see that $\langle J_{y}\rangle_{SS}$ is always
zero in case of $\omega_1=\omega_2$.
Thus we use $\langle J_{x}\rangle_{SS}$ to measure the system's steady-state
coherence \cite{Chou2008}. This is equivalent to measuring coherence
through $\langle c_{1}^{\dagger}c_{2}\rangle$.
According to Eq.(\ref{eq:Jx}), we can divide the steady state quantum
coherence (SSQC) into steady state vacuum coherence (SSVC) $\langle J_{x}^{V}\rangle$
and steady state thermal coherence (SSTC) $\langle J_{x}^{T}\rangle$, i. e.,{
\begin{eqnarray}
\langle J_{x}^{V}\rangle_{\text{SS}} & = &i\, \sum_{i,j=1,2}\lim_{t\rightarrow \infty}\left[C_{i}^{2*}  C_{j}^{1}  -C_{i}^{1*}  C_{j}^{2}  \right]\langle\hat{c}_{i}^{0\,\dagger}\hat{c}_{j}^{0}\rangle,\label{Jxv}\\
\langle J_{x}^{T}\rangle_{\text{SS}}  & = & \int\text{d}\omega\,J(\omega)\mathcal B_x(\omega)\frac{1}{\exp\left(\beta\omega\right)-1}\label{Jxt},
\end{eqnarray}}
{with $\mathcal{B}_{x}=\mathcal{B}_{x}^{c}+\mathcal{B}_{x}^{i}$. Here,
\begin{widetext}
\begin{eqnarray*}
\mathcal{B}_{x}^{c} &=& \tilde{B}_{0,\alpha}^{1*} \tilde{B}_{0,\alpha}^{2} + \tilde{B}_{0,\alpha}^{2*} \tilde{B}_{0,\alpha}^{1}
= \frac{2\,\cos^{4}\theta\,\left( (\omega_1 - \omega)^2 + \gamma^2 \right)}{\left( \Gamma\,\gamma \left( \cos^{4}\theta + \frac{\sin^{4}\theta}{2} \right) - (\omega_1 - \omega)^2 \right)^2 + \gamma^2 (\omega_1 - \omega)^2}, \\
\mathcal{B}_{x}^{i} &=& \sum_{i=1,2} \left( \tilde{B}_{i,\alpha}^{1*} \tilde{B}_{i,\alpha}^{2} + \tilde{B}_{i,\alpha}^{2*} \tilde{B}_{i,\alpha}^{1} \right) \\
&=& -\frac{2\,\Gamma\,\gamma\,\cos^{4}\theta\,\sin^{4}\theta\, \left( \gamma^2 + (\omega_1 - \omega)^2 \right) \left( \frac{\Gamma\,\gamma}{2} \left( \sin^{4}\theta + \cos^{4}\theta \right) - (\omega_1 - \omega)^2 \right)}{\left( \gamma^2 (\omega_1 - \omega)^2 + \left( \frac{\Gamma\,\gamma\,\sin^{4}\theta}{2} - (\omega_1 - \omega)^2 \right)^2 \right) \left( \gamma^2 (\omega_1 - \omega)^2 + \left( \Gamma\,\gamma \left( \frac{\sin^{4}\theta}{2} + \cos^{4}\theta \right) - (\omega_1 - \omega)^2 \right)^2 \right)}.
\end{eqnarray*}
\end{widetext}
which originate from the contributions of collective decoherence and individual decoherence, respectively.}
The vacuum  coherence primarily examines the coherence
induced by quantum vacuum fluctuation noise, while thermal coherence
originates from coherence generated by thermal fluctuation noise.

We focus on the SSVC at first. For the completely collective decoherence, i.e., $\theta=0$, we have
from Eq.(\ref{eq:dA1})
\begin{eqnarray*}
\dot{A}_{j}^{1} & = & -2\int_{0}^{t}\text{d}\tau\,f_{0}(t-\tau)A_{j}^{1}(\tau)\\
\dot{A}_{j}^{2} & = & -2i\,\omega_{1}A_{j}^{2}(t)
\end{eqnarray*}
It is straightforward to observe that \(A_j^2\) is immune
from the collective decoherence.
When $t\rightarrow+\infty$, we have $A_{1}^{1}(+\infty)\rightarrow0$,
i.e., $C_{j}^{1}=-C_{j}^{2}$, and $A_{j}^{2}(t)=\exp(-2i\,\omega_{1}t).$
Thus we have
\begin{eqnarray*}
C_{1}^{1} & = & \frac{\exp(-2i\,\omega_{1}t)}{2},\,C_{2}^{1}  =  -\frac{\exp(-2i\,\omega_{1}t)}{2},\\
C_{1}^{2} & = & -\frac{\exp(-2i\,\omega_{1}t)}{2}\,C_{2}^{2}  =  \frac{\exp(-2i\,\omega_{1}t)}{2}.
\end{eqnarray*}
According to Eq.(\ref{Jxv}), it can be observed that, when the initial state of
oscillators system is $|\psi_{1}\rangle=(|01\rangle-|10\rangle)/\sqrt{2}$,
$|\langle J_{x}^{V}\rangle_{SS}|$ can reach its maximal value, while it is
zero for $|\psi_{1}\rangle=(|01\rangle+|10\rangle)/\sqrt{2}$.
On the other hand, for the partial collective decoherence,
 SSVC will be zero in the
steady state. According to Eq. (7), we know that when \(\theta \neq 0\), all \(A_j^i\) will
decay to zero as time approaches infinity. Consequently, \(C_j^i\) will also approach zero
in the long-time limit. From Eq. (17), it is straightforward to conclude that SSVC
vanishes in the presence of partial decoherence.
This also highlights the extreme dependence of SSVC on
collective decoherence.

\begin{figure}
\includegraphics[scale=0.52]{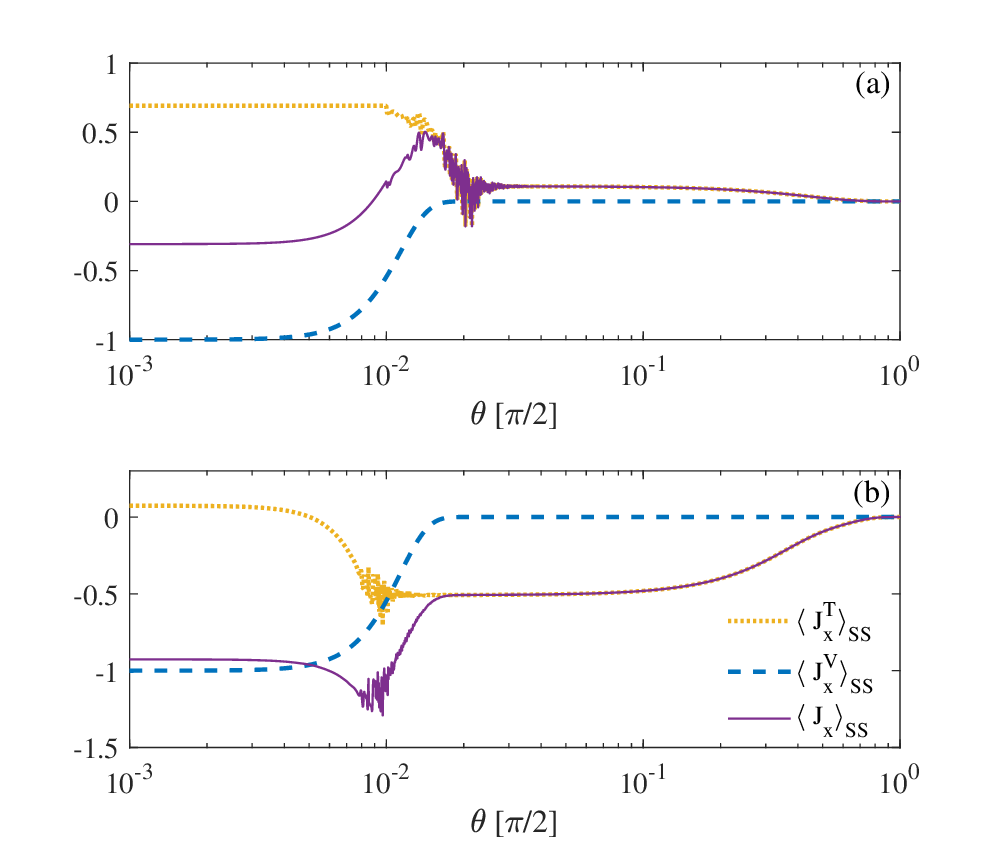}
\caption{{The mean values of the  steady state coherence $\langle J_{x}\rangle$} as a function of $\theta$
(in unit of $\pi/2$) for (a) \(\gamma=0.1\Gamma\) and (b) \(\gamma=10\Gamma\).
The other parameters are chosen as
$\Gamma=\omega_1$, $\omega_2=\omega_0=\omega_1$, $T=\omega_0$.
We set $\omega_1=1$ as a unit for the other parameters.}
\label{fig:theta}
\end{figure}

In Figs. \ref{fig:theta} (a) and (b), we present SSTC as a function of $\theta$
for \(\gamma=0.1\Gamma\) and (b) \(\gamma=10\Gamma\),  respectively. The former is in
the non-markovian regime and the latter is in the markovian regime.
Here, the initial state of the system is chosen as $|\psi_{1}\rangle=\sin\kappa|01\rangle
+\cos\kappa|10\rangle$ with $\kappa=-\pi/4$. It is evident that the dependence of SSTC on
\(\theta\) differs significantly between the Markovian and non-Markovian regimes.
 As \(\theta\) approaches 0, the system's steady state gradually converges to the result of
complete collective decoherence. At this point, we have
\begin{eqnarray}
\lim_{\theta\rightarrow0}\mathcal B_{x}&=&\frac{2\,\left(\left(\omega_{1}-\omega\right)^{2}+\gamma^{2}\right)}
{K(\omega)}\nonumber\\
&-&\frac{\Gamma\,\gamma\,\left(\Gamma\,\gamma-2\,\left(\omega_{1}-\omega\right)^{2}\right)}{\left(\omega_{1}-\omega\right)^{2}
K(\omega)}O(\sin^{4}\theta)\label{eq:mBx}
\end{eqnarray}
with $K(\omega)=\left(\Gamma\,\gamma-\left(\omega_{1}-\omega\right)^{2}\right)^{2}
+\gamma^{2}\left(\omega_{1}-\omega\right)^{2}$, where the first term attributes
to the collective decoherence and the second term comes from the individual decoherence.
 With increasing \(\theta\), SSTC decreases in both
Markovian and non-Markovian cases. However, for the non-Markovian case, SSTC first
decreases to 0 and then gradually increases, while for the Markovian case, SSTC decreases
to 0 and continues to negative values. Eventually, as \(\theta\) approaches \(\pi/2\)
(the regime of complete individual decoherence), SSTC vanishes. Two observations are
noteworthy here: On the one hand, the variation
of SSTC with \(\theta\) is not monotonic, nor does it rapidly approach 0. This is in stark
contrast to results from previous studies \cite{Ablimit2024}. On the other hand, when \(\theta\) is small,
non-Markovian SSTC $|\langle J_{x}^{V}\rangle_{SS}|$ is
stronger than its Markovian counterpart.
Conversely, when \(\theta > \pi/200\), the Markovian SSTC becomes stronger than the
non-Markovian case.

{
Note that in FIG. \ref{fig:theta}, regardless of whether the dynamics are Markovian
or non-Markovian, SSTC exhibits rapid oscillations within a certain range of $\theta$.
These oscillations originate from the divergent behavior in the frequency integration, as evident from Eq. (\ref{eq:mBx}). When $\theta = 0$, only the first term in
 $\mathcal{B}_x$ contributes, corresponding to complete collective decoherence, and no
 divergence occurs. As $\theta$ increases, the second term in Eq. (\ref{eq:mBx}), associated with
 individual decoherence, becomes relevant. This term contains a factor proportional to
 $(\omega_1 - \omega)^{-2}$, which leads to a divergence when integrated over frequency.
 This divergence manifests as rapid oscillations in the numerical results.}

\begin{figure}
\includegraphics[scale=0.5]{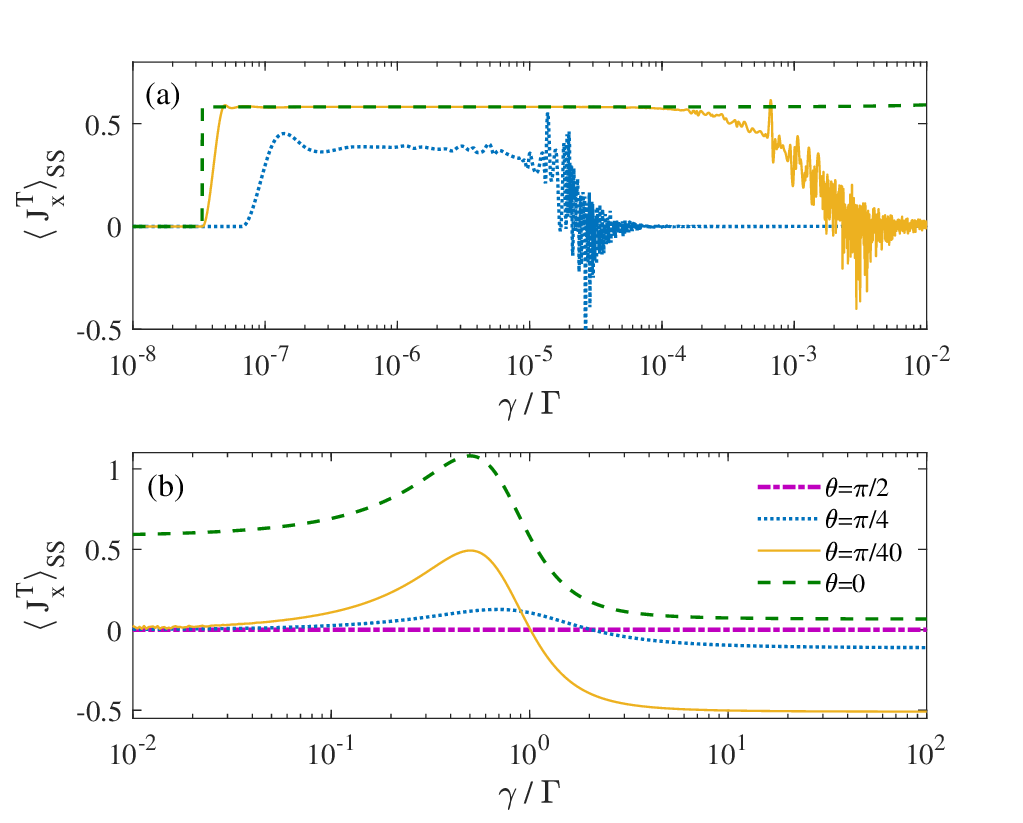}
\caption{{The Steady-state thermal coherence  as a function of the spectral width \(\gamma\)
for different proportions of collective decoherence with (a) \(\gamma/\Gamma \in [10^{-8}, 10^{-2}]\)
and (b) \(\gamma/\Gamma \in [10^{-2}, 10^2]\). The parameters are \(\kappa = -\pi/4\), \(\omega_1 = \omega_2 = \Gamma\),
\(\omega_0 = \Gamma\), and \(T = \omega_0\), with \(\Gamma = 1\) as the unit.} }
\label{fig:gamma}
\end{figure}

When $\gamma/\Gamma$ is less than a particular number, coherence flows back into
the system during the dynamics, demonstrating non-Markovianity. We numerically calculated
the dependence of SSTC on the environmental spectral width \(\gamma\), as shown in Figs. \ref{fig:gamma}.
The results reveal a complex relationship between SSTC and \(\gamma\) under different
proportions of collective decoherence. In Fig. \ref{fig:gamma} (a), we illustrate the behavior of
correlations as
\(\gamma/\Gamma\) approaches 0. The numerical results show that when
\(\gamma/\Gamma < 10^{-8}\), SSTC approaches zero
regardless of the value of \(\theta\). As \(\gamma/\Gamma\) increases, the dynamics exhibit
varying degrees of enhancement under different proportions of collective decoherence.
Subsequently, SSTC reaches its respective peak value and then decays to zero.. This behavior can
be explained by the analytical results. For $\gamma/\Gamma\rightarrow 0$, the spectral
density becomes $J(\omega'-\omega_0)=\frac{\Gamma\gamma}{2}\delta(\omega'-\omega_0)$.
When \(\theta = 0\), individual decoherence does not contribute to the dynamics. In this
case, we have
\[
 \langle J_{x}^{T} \rangle_{SS} = \frac{\gamma}{\Gamma}\frac{1}{\exp(\beta\omega_{1})-1},
\]
indicating that as \(\gamma \to 0\), even under complete collective decoherence, the SSTC
will completely vanish. When \(\theta\neq 0\), it yields
\begin{eqnarray*}
 \lim_{\gamma/\Gamma\rightarrow0}&& \langle J_{x}^{T}\rangle_{SS} =\\
&&\frac{4\,\gamma\cos^{4}\theta}{\Gamma\,\sin^{4}\theta\,
\left(\sin^{4}\theta+2\,\cos^{4}\theta\right)^{2}}\frac{1}{\exp(\beta\omega_{1})-1},
\end{eqnarray*}
which leads to $\langle J_{x}^{T}\rangle_{SS}=0$ with $\gamma=0$.
In the regime where \(\gamma / \Gamma \ll 1\), if \(\gamma / \Gamma \sim \sin^4 \theta\), the SSTC will
exhibit a clear dependence on \(\theta\), as the contribution from individual decoherence
becomes significant and interacts with collective decoherence. However, if
\(\gamma / \Gamma \gg \sin^4 \theta\), the integral governing the SSTC starts to diverge,
due to
\begin{eqnarray*}
 \lim_{\gamma/\Gamma\rightarrow0} \mathcal  B_{x}
 =\frac{2\cos^{4}\theta}{\left(\omega_{1}-\omega\right)^{2}}.
\end{eqnarray*}
This divergence continues until the environmental spectral width \(\gamma\) grows sufficiently
large, at which point the environment can not be considered as a single mode bosonic reservoir,
leading to a decay  in SSTC as \(\gamma\) increases further. These behaviors
occur only in the region where \(\gamma/\Gamma < 0.01\), corresponding to the non-Markovian
dynamics regime.

As \(\gamma / \Gamma\) continues to increase, the system gradually transitions into the Markovian
regime. {As shown in Fig. \ref{fig:gamma} (b), for both the case of
complete and partial collective  decoherence, SSTC first increases and then decreases. }
SSTC reaches its peak at the critical point between the Markovian and non-Markovian regimes and
subsequently decreases, eventually becoming negative. Since the measure of coherence depends on
the absolute value of \(\langle J_x^T \rangle_{SS}\), SSTC vanishes entirely at certain specific values of
\(\gamma / \Gamma\). In the Markovian limit, i.e.,$\gamma/\Gamma\rightarrow\infty$, it is easy to verify
\[
J(\omega)=\lim_{\gamma/\Gamma\rightarrow\infty}\frac{\Gamma \gamma^2/2\pi}{(\omega-\Omega)^2+\gamma^2}=\frac{\Gamma}{2\pi},
\]
and
\begin{eqnarray*}
&&\lim_{\gamma/\Gamma\rightarrow \infty}\mathcal B_x=\frac{2\,\pi\,\cos^{4}\theta}{\Gamma\left(\cos^{4}\theta+\frac{\sin^{4}\theta}{2}\right)}L_2(\omega)\\
&&-\frac{\pi\,\Gamma\,\cos^{4}\theta\,\sin^{4}\theta\,\left(\sin^{4}\theta+\cos^{4}\theta\right)}{\left(\frac{\sin^{4}\theta}{2}+\cos^{4}\theta\right)
\left(\left(\omega_{1}-\omega\right)^{2}+\left(\frac{\Gamma\,\sin^{4}\theta}{2}\right)^{2}\right)}L_2(\omega),
\end{eqnarray*}
where
\[L_2(\omega)=\frac{\Gamma\,\left(\cos^{4}\theta+\frac{\sin^{4}\theta}{2}\right)/\pi}{\Gamma^{2}
\,\left(\cos^{4}\theta+\frac{\sin^{4}\theta}{2}\right)^{2}+\left(\omega_{1}-\omega_{\alpha}\right)^{2}}\]
 is a Lorentzian-type function. In the Markovian limit, i.e., \(\Gamma / \gamma \to 0\), \(L_2 = \delta(\omega - \omega_1)\),
 and consequently,
\[
\langle J_x^T \rangle_{SS} = \frac{-2\cos^4\theta}{\sin^4\theta} \frac{1}{\exp(\beta \omega_1) - 1}.
\]
The analytical results for \(\theta \neq 0\) are consistent with the numerical results shown in Fig.
\ref{fig:gamma} (b). For the case of complete collective decoherence, individual decoherence does
not contribute to the system's steady-state behavior, and the second term in the equation vanishes.
In this case, we obtain
\[
\langle J_x^T \rangle_{SS} = \frac{1}{\exp(\beta \omega_1) - 1},
\]
indicating that the oscillator system interacts with the common thermal bath and ultimately
reaches the corresponding thermal equilibrium state.

\begin{figure}
\includegraphics[scale=0.55]{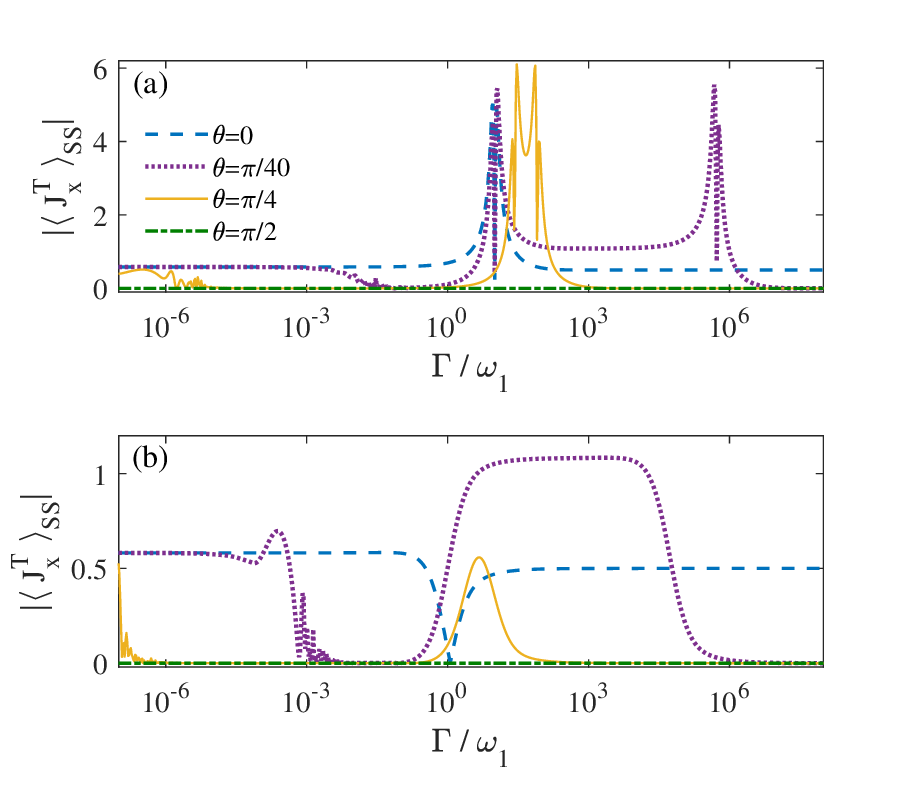}
\caption{The steady state quantum coherences as a function of the coupling
strength $\Gamma$ for (a) $\gamma=0.1\Gamma$ and (b) $\gamma=10\Gamma$
with different proportions of collective decoherence $\theta$.
The other parameters are chosen as $\kappa=-\pi/4$,
$\omega_0=\omega_2=\omega_1$,  and $T=\omega_0$.
We set $\omega_1=1$ as a unit for the other parameters.}

\label{fig:Gamma}\end{figure}

In Fig. \ref{fig:Gamma}, we plot the variation of SSTC as a
function of the coupling strength \(\Gamma\)
under different proportions of collective decoherence. We consider two cases:
\(\gamma = 0.1\Gamma\) and \(\gamma = 10\Gamma\), corresponding
to non-Markovian and Markovian dynamics, respectively.
Notably, according to Eq. (18), in the limit \(\Gamma \to 0\), we have
\[
\lim_{\Gamma\rightarrow 0} \mathcal{B}_{x} \propto (\omega - \omega_1)^{-2}.
\]
This indicates that the integration over \(\omega_\alpha\) diverges as \(\Gamma \to 0\).
However, if \(\Gamma = 0\), we unambiguously obtain \(\langle J_x^T \rangle_{SS} = 0\),
due to \(J(\omega_\alpha)=0\).
For the case of complete collective decoherence (\(\theta = 0\)), as shown by the blue dashed
lines in Figs. \ref{fig:Gamma}(a) and (b), SSTC exhibits identical behavior in both the weak
and strong coupling regimes. {An interesting difference emerges for  \(\Gamma \sim 10\omega_1\). The analytical solution uniquely captures the expected multi-peak structure induced by $\Gamma$ in the non-Markovian regimes . This indicates that conventional master equations, which rely on weak coupling  assumptions, fail to describe the behavior of open systems in the strong-coupling (large-$\Gamma$) regime~\cite{Ablimit2024}. This highlights the importance of exactly
solvable models, which provide an accurate description of both the dynamics and
the steady-state properties across all coupling strengths.
In contrast, approaches such as reaction-coordinate mapping or perturbative
treatments inevitably rely on approximations, and therefore cannot fully account
for the rich behaviors that emerge beyond their validity regimes.
 }For the case of partial collective decoherence,
we find that regardless of whether the dynamics are Markovian or non-Markovian, there
exists a range of \(\Gamma\) where SSTC remains nonzero. As \(\theta\) increases,
the region where SSTC is nonzero gradually shrinks until it eventually disappears.
From Fig. \ref{fig:Gamma} (a), we observe that in the non-Markovian regime,
by choosing an appropriate coupling strength \(\Gamma\), SSTC can be stronger
than in the case of complete collective decoherence. However, in the Markovian regime,
this phenomenon only occurs when the proportion of collective decoherence is dominant,
i.e., when \(\theta\) is very small. Finally, as \(\Gamma\) approaches infinity,
SSTC vanishes. This can be attributed to strong coupling with individual environments,
which significantly suppresses the coherence induced by collective decoherence.
In the ultra-strong individual decoherence regime, the coherence effects originating
from collective decoherence are effectively erased.

\begin{figure}
\includegraphics[scale=0.5]{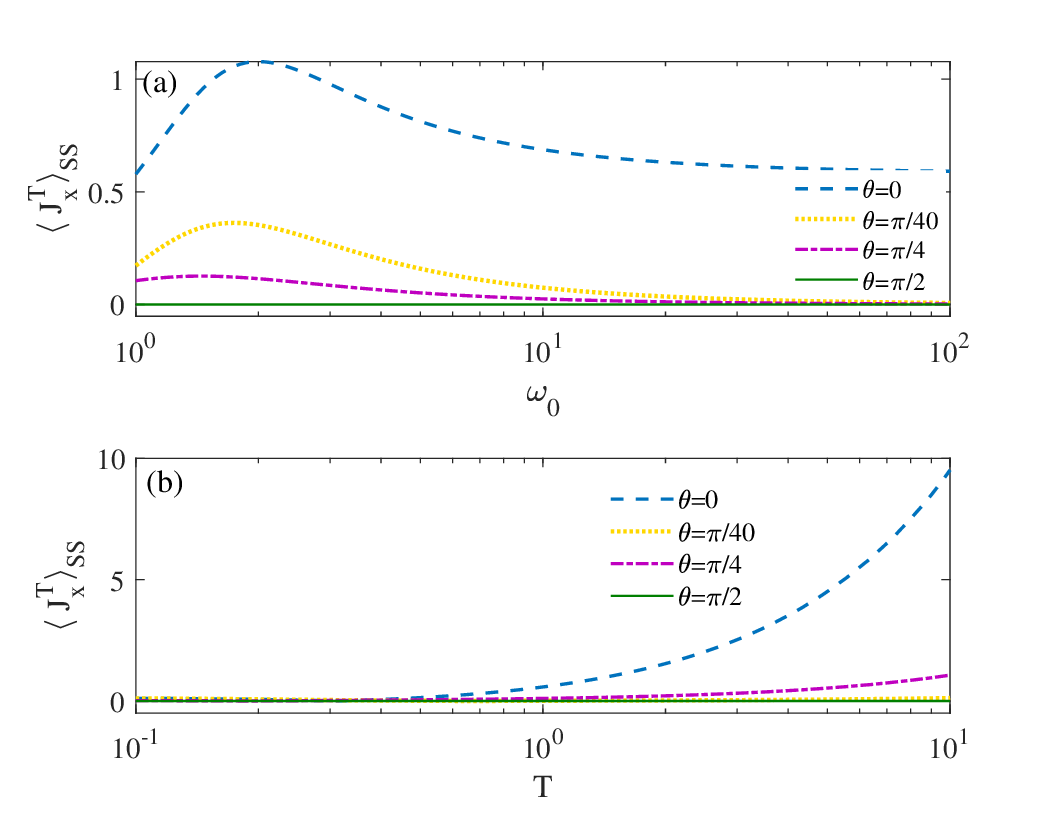}
\caption{The steady state quantum coherences as a function of (a) reservoir's
center frequency $\omega_0$ for $T=\omega_1$ and
(b) environmental temperature  $T$ for $\omega_0=\gamma$
with  different proportions of collective decoherence $\theta$.
The other parameters are chosen as $\kappa=-\pi/4$,
$\gamma=\omega_2=\omega_1$,  and
We set $\omega_1=1$ as a unit for the other parameters.}
\label{fig:GT}
\end{figure}

In Fig. \ref{fig:GT} (a), we plot the steady-state thermal coherence
as a function of the center frequency of the reservior \(\omega_0\) for different values of
\(\theta\). For the case of complete decoherence, initially, SSTC exhibits a
significant enhancement as \(\omega_0\) increases, followed by a gradual decrease
(blue dashed line). However, even in the regime where \(\omega_0 \gg \omega_1\),
SSTC remains observable. In contrast, when individual decoherence is involved,
although the overall trend of SSTC with respect to \(\omega_0\) remains unchanged,
SSTC completely vanishes in the \(\omega_0 \gg \omega_1\) regime.
Similarly, as \(\theta\) increases, the observable SSTC weakens significantly.
This suggests that appropriately manipulating the detuning between the system and the
reservoir can help maintain stronger steady-state coherence.
We also considered the impact
of environmental temperature on the SSQC (see Fig. \ref{fig:GT} (b)). As expected, the
SSQC exhibits an increasing trend with rising temperature. It is evident that
this dependence arises only in the presence of individual decoherence.
For the case of complete collective decoherence, the SSQC first decreases
with temperature and then increases again. The initial decrease is primarily
due to thermal correlations offsetting part of the vacuum correlations at lower
temperatures. Once the thermal correlations surpass the vacuum correlations,
the SSQC begins to grow with temperature. It is worth noting that significant
thermal correlations cannot be observed in practice. This is mainly because the
temperature dependence of the SSQC  is highly sensitive to individual decoherence.
For instance, when \(\theta = \pi/16\), we observe almost no variation in thermal
correlations with temperature (green dotted line).

\begin{figure}
\includegraphics[scale=0.45]{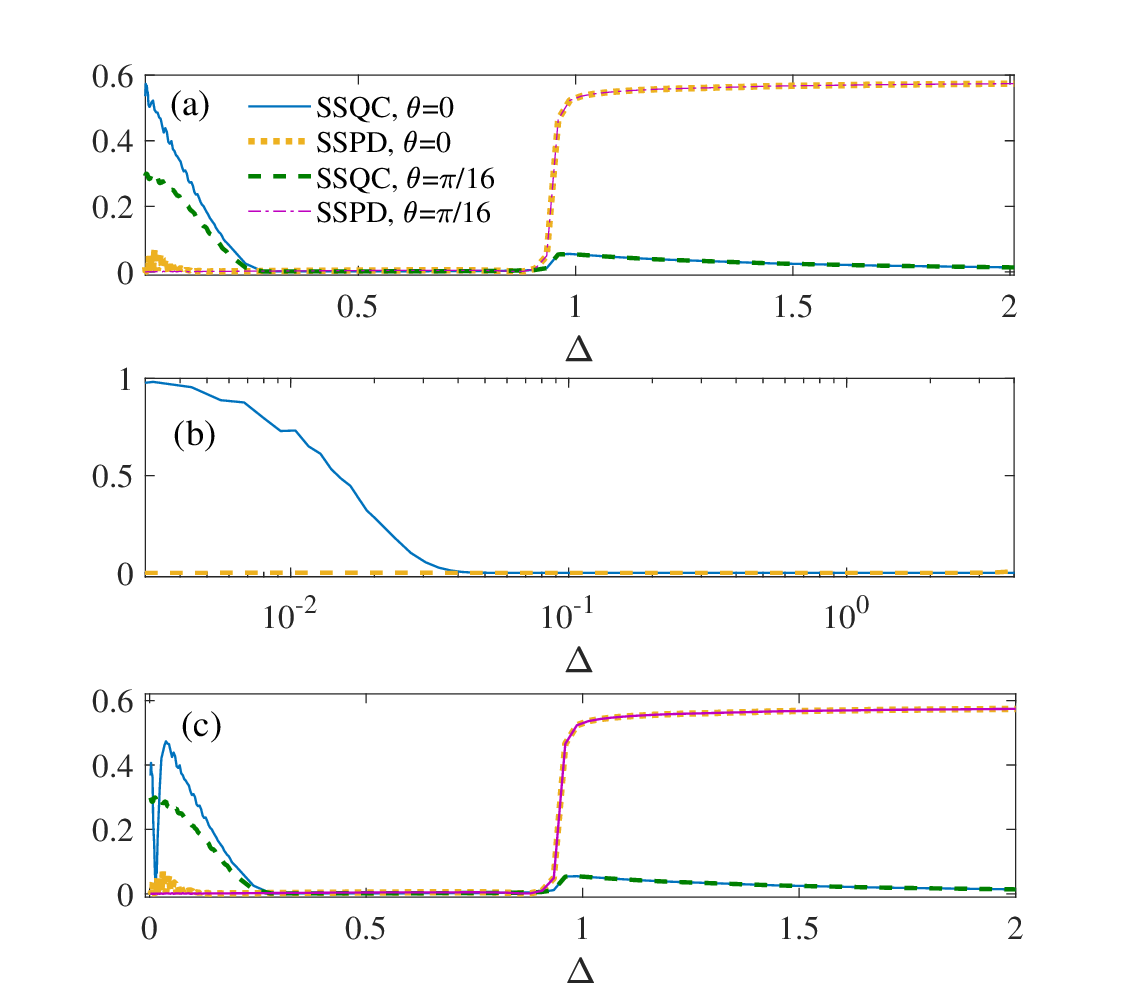}
\caption{The steady state coherences $|J_x|$
and the steady state population differences $|J_z|$
as a function of frequency difference $\Delta$ from (a) thermal,
(b) dissipative, and (c) total difference. The other parameters are chosen as $\kappa=-\pi/4$,
$\gamma=0.1\Gamma$, $\omega_1=100\Gamma$, $\omega_0=100\Gamma$, and $T=\omega_0$.
We set $\Gamma=1$ as a unit for the other parameters.}
\label{fig:DT}
\end{figure}

Finally, we consider the case where \(\omega_1 \neq \omega_2\).
We present the variations of quantum coherence and steady-state population
difference (SSPD) as a function of the frequency difference
\(\Delta = \omega_1 - \omega_2\). To clearly illustrate the results, we have
plotted thermal coherence, vacuum coherence, and quantum coherence
in Figs. \ref{fig:DT} (a), (b), and (c), respectively. As \(\Delta\) increases, both thermal
coherence and vacuum coherence exhibit rapid decay. However, it is evident
that vacuum coherence decays faster than thermal coherence (blue solid line).
When \(\Delta\) reaches a certain value, both SSQC and SSPD undergo a
sudden increase. The steady-state thermal coherence and steady-state
population difference experience a noticeable change. As \(\Delta\)
continues to grow, both coherence and population difference approach
stable values.

\section{Conclusion} \label{Conc}

In this work, we have studied the steady-state coherence of a system comprising
two non-interacting harmonic oscillators coupled to both individual and collective
environments. By introducing a tunable parameter that adjusts the proportions of collective
to individual decoherence, we derived exact dynamical equations and analytical
solutions in the Heisenberg picture. This approach enabled us to precisely quantify
the system's steady-state coherence and investigate the effects of both types of
decoherence on the system's behavior.

Our findings show that steady-state coherence is highly sensitive to the balance
between collective and individual decoherence. In the case of perfect collective
decoherence, the steady-state coherence depends on the system's initial state,
whereas in the presence of partial collective decoherence, this dependence is
suppressed. Furthermore, contrary to previous studies \cite{Ablimit2024},
we found that strong
non-Markovianity does not induce steady-state coherence. These results highlight
the importance of considering both collective and individual decoherence in realistic
models of open quantum systems, as even a small contribution of individual
decoherence can significantly impact the system's coherence.


{While this work focuses on analytical dynamical approach
, some of the results here can
also be interpreted from the perspective of the Hamiltonian
of mean force (HMF), a framework that characterizes
equilibrium properties of strongly coupled open systems~\cite{Strasberg2016,Hartmann2020,Weiss2021} .
This alternative viewpoint suggests that
the asymptotic state reached by the system corresponds
to a reduced Gibbs state governed by an effective HMF,
incorporating renormalization effects due to the environments.
Exploring this connection in detail would be an
interesting direction for future work. Another promising strategy is the reaction coordinate  (RC) mapping, which extracts one (or a few) environmental modes after a transformation from the spectral density and elevates them to the extended system~\cite{RevModPhys.92.041002,Onsager1933,Campisi2009,Hilt2011} . The remaining bath is redefined and can be approximated as a memoryless white noise environment, corresponding to the Markovian approximation of the residual bath. It can be considered a relatively general method. Yet our analytical solutions of specific bilinear models serve as benchmarks for all methods, including the RC and HMF, when they are applied to the same bilinear Bosonic models.}

{Given the simulation results and the trends identified, we note that the proposed model
has strong experimental feasibility in several state-of-the-art platforms. In particular,
superconducting circuit quantum electrodynamics  systems allow precise
engineering of dissipation channels, enabling independent and collective reservoir
configurations with controllable strengths. Recent experiments have demonstrated
autonomous entanglement stabilization using engineered lossy resonators \cite{Brown2022}, as well as
hardware-efficient stabilization of entangled states via engineered dissipative channels \cite{Hatridge2024},
all at millikelvin temperatures with tunable spectral properties. Similar capabilities have also
been demonstrated in optomechanical systems, where mechanical resonators coupled to
optical cavities enable precise control over system-bath interactions and allow the observation
of synchronized optomechanical squeezing \cite{Molinares2025}. Therefore, the key findings of this work---
including the sensitivity of steady-state coherence to the admixture of individual decoherence
and its dependence on environmental parameters---can be directly tested in these platforms.
Our results thus offer both theoretical insights and concrete guidance for experimental studies
aiming to observe and control steady-state quantum coherence.}

Our work provides a deeper understanding of how environmental interactions
influence the steady-state properties of quantum systems and offers valuable
insights into controlling decoherence in experimental setups. The findings
pave the way for future research aimed at optimizing the coherence of quantum
systems, with potential applications in quantum computing, quantum metrology,
and other quantum technologies.

\section*{Acknowledgement}

This work was supported by the National Natural Science Foundation of China (NSFC)
under Grants No. 12205037, 12375009, 12075050; by the Fundamental Research Funds for the
Central Universities under Grant No. 04442024072; by the grant PID2021-126273NB-I00 funded by
MCIN/AEI/10.13039/501100011033, and by "ERDF A way of making Europe" and the
Basque Government through grant number IT1470-22.  Professor Brumer's work was supported by
NSERC Canada.

\appendix

\begin{widetext}

\section{Detailed Derivation for $D_{j,\alpha}$}\label{Dij}

For thermal noise, the collective decoherence satisfies
\begin{eqnarray*}
\dot{B}_{0,\alpha}^{1}(t) & = & -i\,\omega_{1}B_{0,\alpha}^{1}(t)-i\,\cos^2\theta\,g_{\alpha}^{*}\exp\left(-i\:\omega_{0,\alpha}t\right)-\sin^{4}\theta\int_{0}^{t}\text{d}\tau\,f(t-\tau)B_{0,\alpha}^{1}(\tau)-\cos^{4}\theta\int_{0}^{t}\text{d}\tau\,f(t-\tau)\sum_{p=1,2}B_{0,\alpha}^{p}(\tau),\\
\dot{B}_{0,\alpha}^{2}(t) & = & -i\,\omega_{1}B_{0,\alpha}^{2}(t)-i\,\cos^2\theta\,g_{\alpha}^{*}\exp\left(-i\:\omega_{0,\alpha}t\right)-\sin^{4}\theta\int_{0}^{t}\text{d}\tau\,f(t-\tau)B_{0,\alpha}^{2}(\tau)-\cos^{4}\theta\int_{0}^{t}\text{d}\tau\,f(t-\tau)\sum_{p=1,2}B_{0,\alpha}^{p}(\tau).
\end{eqnarray*}
It is convenient to define
\begin{eqnarray*}
D_{0,\alpha}^{1} & = & B_{0,\alpha}^{1}+B_{0,\alpha}^{2},\\
D_{0,\alpha}^{2} & = & B_{0,\alpha}^{1}-B_{0,\alpha}^{2},
\end{eqnarray*}
which fulfill
\begin{eqnarray*}
\dot{D}_{0,\alpha}^{1}(t) & = & -i\,\omega_{1}D_{0,\alpha}^{1}(t)-2i\,\cos^2\theta\,g_{\alpha}^{*}\exp\left(-i\:\omega_{0,\alpha}t\right)-\left(\sin^{4}\theta+2\cos^{4}\theta\right)\int_{0}^{t}\text{d}\tau\,f(t-\tau)D_{0,\alpha}^{1}(\tau),\\
\dot{D}_{0,\alpha}^{2}(t) & = & -i\,\omega_{1}D_{0,\alpha}^{2}(t)-\sin^{4}\theta\int_{0}^{t}\text{d}\tau\,f(t-\tau)D_{0,\alpha}^{2}(\tau),
\end{eqnarray*}
Differentiating the above equation again, we have
\begin{eqnarray*}
\ddot{D}_{0,\alpha}^{1}(t) & = & -i\,\omega_{1}\dot{D}_{0,\alpha}^{1}(t)-2\:\omega_{0,\alpha}\,\cos^2\theta\,g_{\alpha}^{*}\exp\left(-i\:\omega_{0,\alpha}t\right)-\left(\sin^{4}\theta+2\cos^{4}\theta\right)f(0)D_{0,\alpha}^{1}\\
 &  & +\left(\sin^{4}\theta+2\cos^{4}\theta\right)\left(\gamma+i\,\omega_{0}\right)\int_{0}^{t}\text{d}\tau\,f(t-s)D_{0,\alpha}^{1}(\tau),\\
\ddot{D}_{0,\alpha}^{2}(t) & = & -i\,\omega_{1}\dot{D}_{0,\alpha}^{2}(t)-\sin^{4}\theta f(0)D_{0,\alpha}^{2}(\tau)+\sin^{4}\theta\left(\gamma+i\,\omega_{0}\right)\int_{0}^{t}\text{d}\tau\,f(t-s)D_{0,\alpha}^{2}(\tau).
\end{eqnarray*}
Substituting
\begin{eqnarray*}
\int_{0}^{t}\text{d}\tau\,f(t-\tau)D_{0,\alpha}^{1}(\tau) & = & -\frac{\dot{D}_{0,\alpha}^{1}(t)+i\,\omega_{1}D_{0,\alpha}^{1}(t)+2i\,\cos^2\theta\,g_{\alpha}^{*}\exp\left(-i\:\omega_{0,\alpha}t\right)}{\sin^{4}\theta+2\cos^{4}\theta},\\
\int_{0}^{t}\text{d}\tau\,f(t-\tau)D_{0,\alpha}^{2}(\tau) & = & -\frac{\dot{D}_{0,\alpha}^{2}(t)+i\,\omega_{1}D_{0,\alpha}^{2}(t)}{\sin^{4}\theta}.
\end{eqnarray*}
into above equation, we have
\begin{eqnarray*}
\ddot{D}_{0,\alpha}^{1}(t) & = & -i\,\omega_{1}\dot{D}_{0,\alpha}^{1}(t)-2\:\omega_{0,\alpha}\,\cos^2\theta\,g_{\alpha}^{*}\exp\left(-i\:\omega_{0,\alpha}t\right)
-\left(\sin^{4}\theta+2\cos^{4}\theta\right)\frac{\Gamma\,\gamma}{2}D_{0,\alpha}^{1}\\
 &  & -\left(\gamma+i\,\omega_{0}\right)\dot{D}_{0,\alpha}^{1}(t)-i\,\omega_{1}\left(\gamma+i\,\omega_{0}\right)D_{0,\alpha}^{1}(t)
 -2i\,\cos^2\theta\,g_{\alpha}^{*}\left(\gamma+i\,\omega_{0}\right)\exp\left(-i\:\omega_{0,\alpha}t\right)\\
 & = & -\left(\gamma+i\,\left(\omega_{0}+\omega_{1}\right)\right)\dot{D}_{0,\alpha}^{1}(t)-\left(\left(\left(\sin^{4}\theta+2\cos^{4}\theta\right)
 \frac{\Gamma\,\gamma}{2}-\omega_{1}\omega_{0}\right)+i\,\omega_{1}\gamma\right)D_{0,\alpha}^{1}(t)\\
 &  & -2\:\cos^2\theta\,g_{\alpha}^{*}\left(\left(\omega_{0,\alpha}-\omega_{0}\right)+i\,\gamma\right)\exp\left(-i\:\omega_{0,\alpha}t\right),\\
\ddot{D}_{0,\alpha}^{2}(t) & = & -\left(i\,\left(\omega_{1}+\omega_{0}\right)+\gamma\right)\dot{D}_{0,\alpha}^{2}(t)-\left(\left(\sin^{4}\theta\frac{\Gamma\,\gamma}{2}-\omega_{0}\omega_{1}\right)
+i\,\gamma\omega_{1}\right)D_{0,\alpha}^{2}(t).
\end{eqnarray*}
The individual decoherence satisfies
\begin{eqnarray*}
\dot{B}_{1,\alpha}^{1}(t) & = & -i\,\omega_{1}B_{1,\alpha}^{1}(t)-\sin^{4}\theta\int_{0}^{t}\text{d}\tau\,f(t-\tau)B_{1,\alpha}^{1}(\tau)-\cos^{4}\theta\int_{0}^{t}\text{d}\tau\,f(t-\tau)\sum_{p=1,2}B_{1,\alpha}^{p}(\tau)\\
 &  & -i\,\sin^2\theta\,g_{\alpha}^{*}\exp\left(-i\:\omega_{1,\alpha}t\right),\\
\dot{B}_{1,\alpha}^{2}(t) & = & -i\,\omega_{1}B_{1,\alpha}^{2}(t)-\sin^{4}\theta\int_{0}^{t}\text{d}\tau\,f(t-\tau)B_{1,\alpha}^{2}(\tau)-\cos^{4}\theta\int_{0}^{t}\text{d}\tau\,f(t-\tau)\sum_{p=1,2}B_{1,\alpha}^{p}(\tau).
\end{eqnarray*}
By defining
\begin{eqnarray*}
D_{1,\alpha}^{1} & = & B_{1,\alpha}^{1}+B_{1,\alpha}^{2},\\
D_{1,\alpha}^{2} & = & B_{1,\alpha}^{1}-B_{1,\alpha}^{2},
\end{eqnarray*}
it yields
\begin{eqnarray*}
\dot{D}_{1,\alpha}^{1}(t) & = & -i\,\omega_{1}D_{1,\alpha}^{1}(t)-\left(\sin^{4}\theta+2\cos^{4}\theta\right)\int_{0}^{t}\text{d}\tau\,f(t-\tau)D_{1,\alpha}^{1}(\tau)-i\,\sin^2\theta\,g_{\alpha}^{*}\exp\left(-i\:\omega_{1,\alpha}t\right),\\
\dot{D}_{1,\alpha}^{2}(t) & = & -i\,\omega_{1}D_{1,\alpha}^{2}(t)-\sin^{4}\theta\int_{0}^{t}\text{d}\tau\,f(t-\tau)D_{1,\alpha}^{2}(\tau)-i\,\sin^2\theta\,g_{\alpha}^{*}\exp\left(-i\:\omega_{1,\alpha}t\right).
\end{eqnarray*}
Differentiating the above equation again, we have
\begin{eqnarray*}
\ddot{D}_{1,\alpha}^{1}(t) & = & -i\,\omega_{1}\dot{D}_{1,\alpha}^{1}(t)-\left(\sin^{4}\theta+2\cos^{4}\theta\right)f(0)D_{1,\alpha}^{1}(t)-\left(\sin^{4}\theta+2\cos^{4}\theta\right)
\int_{0}^{t}\text{d}\tau\,\dot{f}(t-\tau)D_{1,\alpha}^{1}(\tau)\nonumber\\
&&-\omega_{1,\alpha}\,\sin^2\theta\,g_{\alpha}^{*}\exp\left(-i\:\omega_{1,\alpha}t\right),\\
\ddot{D}_{1,\alpha}^{2}(t) & = & -i\,\omega_{1}\dot{D}_{1,\alpha}^{2}(t)-\sin^{4}\theta f(0)D_{1,\alpha}^{2}(t)+-\sin^{4}\theta\int_{0}^{t}\text{d}\tau\,\dot{f}(t-\tau)D_{1,\alpha}^{2}(\tau)\nonumber\\
&&-\omega_{1,\alpha}\,\sin^2\theta\,g_{\alpha}^{*}\exp\left(-i\:\omega_{1,\alpha}t\right).
\end{eqnarray*}
By using
\begin{eqnarray*}
\int_{0}^{t}\text{d}\tau\,f(t-\tau)D_{1,\alpha}^{1}(\tau) & = & -\frac{\dot{D}_{1,\alpha}^{1}(t)+i\,\omega_{1}D_{1,\alpha}^{1}(t)+i\,\sin^2\theta\,g_{\alpha}^{*}\exp\left(-i\:\omega_{1,\alpha}t\right)}{\sin^{4}\theta+2\cos^{4}\theta},\\
\int_{0}^{t}\text{d}\tau\,f(t-\tau)D_{1,\alpha}^{1}(\tau) & = & -\frac{\dot{D}_{1,\alpha}^{2}(t)+i\,\omega_{1}D_{1,\alpha}^{1}(t)+i\,\sin^2\theta\,g_{\alpha}^{*}\exp\left(-i\:\omega_{1,\alpha}t\right)}{\sin^{4}\theta},
\end{eqnarray*}
we have
\begin{eqnarray*}
\ddot{D}_{1,\alpha}^{1}(t) & = & -\left(\gamma+i\,\left(\omega_{0}+\omega_{1}\right)\right)\dot{D}_{1,\alpha}^{1}(t)-\left(i\,\gamma\omega_{1}+\left(\left(\sin^{4}\theta+2\cos^{4}\theta\right)
\frac{\Gamma\,\gamma}{2}-\omega_{0}\omega_{1}\right)\right)D_{1,\alpha}^{1}(t)\\
 &  & -\left(i\,\gamma+\left(\omega_{1,\alpha}-\omega_{0}\right)\right)\,\sin^2\theta\,g_{\alpha}^{*}\exp\left(-i\:\omega_{1,\alpha}t\right),\\
\ddot{D}_{1,\alpha}^{2}(t) & = & -\left(\gamma+i\,\left(\omega_{0}+\omega_{1}\right)\right)\dot{D}_{1,\alpha}^{2}(t)-\left(i\,\gamma\omega_{1}+\left(\sin^{4}\theta
\frac{\Gamma\,\gamma}{2}-\omega_{0}\omega_{1}\right)\right)D_{1,\alpha}^{1}(t)\\
 &  & -\left(i\,\gamma+\left(\omega_{1,\alpha}-\omega_{0}\right)\right)\,\sin^2\theta\,g_{\alpha}^{*}\exp\left(-i\:\omega_{1,\alpha}t\right).
\end{eqnarray*}
At last, we check
\begin{eqnarray*}
\dot{B}_{2,\alpha}^{1}(t) & = & -i\,\omega_{1}B_{2,\alpha}^{1}(t)-\sin^{4}\theta\int_{0}^{t}\text{d}\tau\,f(t-\tau)B_{2,\alpha}^{1}(\tau)-\cos^{4}\theta\int_{0}^{t}\text{d}\tau\,f(t-\tau)\sum_{p=1,2}B_{2,\alpha}^{p}(\tau),\\
\dot{B}_{2,\alpha}^{2}(t) & = & -i\,\omega_{1}B_{2,\alpha}^{2}(t)-\sin^{4}\theta\int_{0}^{t}\text{d}\tau\,f(t-\tau)B_{2,\alpha}^{2}(\tau)-\cos^{4}\theta\int_{0}^{t}\text{d}\tau\,f(t-\tau)\sum_{p=1,2}B_{2,\alpha}^{p}(\tau)\\
 &  & -i\,\sin^2\theta\,g_{\alpha}^{*}\exp\left(-i\:\omega_{2,\alpha}t\right).
\end{eqnarray*}
We define
\begin{eqnarray*}
D_{2,\alpha}^{1} & = & B_{2,\alpha}^{1}+B_{2,\alpha}^{2},\\
D_{2,\alpha}^{2} & = & B_{2,\alpha}^{1}-B_{2,\alpha}^{2},
\end{eqnarray*}
which results in
\begin{eqnarray*}
\dot{D}_{2,\alpha}^{1}(t) & = & -i\,\omega_{1}D_{2,\alpha}^{1}(t)-\left(\sin^{4}\theta+2\cos^{4}\theta\right)\int_{0}^{t}\text{d}\tau\,f(t-\tau)D_{2,\alpha}^{1}(\tau)-i\,\sin^2\theta\,g_{\alpha}^{*}\exp\left(-i\:\omega_{2,\alpha}t\right),\\
\dot{D}_{2,\alpha}^{2}(t) & = & -i\,\omega_{1}D_{2,\alpha}^{2}(t)-\sin^{4}\theta\int_{0}^{t}\text{d}\tau\,f(t-\tau)D_{2,\alpha}^{2}(\tau)+i\,\sin^2\theta\,g_{\alpha}^{*}\exp\left(-i\:\omega_{2,\alpha}t\right).
\end{eqnarray*}
Differentiating the above equation yields
\begin{eqnarray*}
\ddot{D}_{2,\alpha}^{1}(t) & = & -i\,\omega_{1}\dot{D}_{2,\alpha}^{1}(t)-\left(\sin^{4}\theta+2\cos^{4}\theta\right)\left(f(0)D_{2,\alpha}^{1}(t)+\int_{0}^{t}\text{d}\tau\,\dot{f}(t-\tau)D_{2,\alpha}^{1}(\tau)\right)\\
 &  & -\omega_{2,\alpha}\,\sin^2\theta\,g_{\alpha}^{*}\exp\left(-i\:\omega_{2,\alpha}t\right),\\
\ddot{D}_{2,\alpha}^{2}(t) & = & -i\,\omega_{1}\dot{D}_{2,\alpha}^{2}(t)-\sin^{4}\theta\left(f(0)D_{2,\alpha}^{1}(t)+\int_{0}^{t}\text{d}\tau\,\dot{f}(t-\tau)D_{2,\alpha}^{1}(\tau)\right)\\
 &  & +\omega_{2,\alpha}\,\sin^2\theta\,g_{\alpha}^{*}\exp\left(-i\:\omega_{2,\alpha}t\right).
\end{eqnarray*}
By using
\begin{eqnarray*}
\int_{0}^{t}\text{d}\tau\,f(t-\tau)D_{2,\alpha}^{1}(\tau) & = & -\frac{\dot{D}_{2,\alpha}^{1}(t)+i\,\omega_{1}D_{2,\alpha}^{1}(t)+i\,\sin^2\theta\,g_{\alpha}^{*}\exp\left(-i\:\omega_{2,\alpha}t\right)}{\sin^{4}\theta+2\cos^{4}\theta},\\
\int_{0}^{t}\text{d}\tau\,f(t-\tau)D_{2,\alpha}^{2}(\tau) & = & -\frac{\dot{D}_{2,\alpha}^{2}(t)+i\,\omega_{1}D_{2,\alpha}^{2}(t)-i\,\sin^2\theta\,g_{\alpha}^{*}\exp\left(-i\:\omega_{2,\alpha}t\right)}{\sin^{4}\theta},
\end{eqnarray*}
we have
\begin{eqnarray*}
\ddot{D}_{2,\alpha}^{1}(t) & = & -\left(\gamma+i\,\left(\omega_{0}+\omega_{1}\right)\right)\dot{D}_{2,\alpha}^{1}(t)-\left(i\gamma\,\omega_{1}+\left(\sin^{4}\theta+2\cos^{4}\theta\right)
\frac{\Gamma\,\gamma}{2}-\omega_{0}\,\omega_{1}\right)D_{2,\alpha}^{1}(t)\\
 &  & -\left(i\gamma+\omega_{2,\alpha}-\omega_{0}\right)\,\sin^2\theta\,g_{\alpha}^{*}\exp\left(-i\:\omega_{2,\alpha}t\right),\\
\ddot{D}_{2,\alpha}^{2}(t) & = & -\left(\gamma+i\,\left(\omega_{0}+\omega_{1}\right)\right)\dot{D}_{2,\alpha}^{2}(t)-\left(i\gamma\omega_{1}+\sin^{4}\theta
\frac{\Gamma\,\gamma}{2}-\omega_{0}\omega_{1}\right)\,D_{2,\alpha}^{2}(t)\\
 &  & +\left(i\gamma+\left(\omega_{2,\alpha}-\omega_{0}\right)\right)\,\sin^2\theta\,g_{\alpha}^{*}\exp\left(-i\:\omega_{2,\alpha}t\right).
\end{eqnarray*}

\section{Laplace Transform Method}  \label{Lap}

According to Eqs. (\ref{eq:Cij}) and (\ref{eq:Bij}) , we have
\begin{eqnarray*}
\dot{C}_{1}^{j}(t) & = & -i\,\omega_{j}C_{1}^{j}(t)-\int_{0}^{t}\text{d}\tau\,f_{j}(t-\tau)C_{1}^{j}(\tau)-\int_{0}^{t}\text{d}\tau\,f_{0}(t-\tau)\sum_{p=1,2}C_{1}^{p}(\tau),\\
\dot{C}_{2}^{j}(t) & = & -i\,\omega_{j}C_{2}^{j}(t)-\int_{0}^{t}\text{d}\tau\,f_{j}(t-\tau)C_{2}^{j}(\tau)-\int_{0}^{t}\text{d}\tau\,f_{0}(t-\tau)\sum_{p=1,2}C_{2}^{p}(\tau),\\
\dot{B}_{0,\alpha}^{j}(t) & = & -i\,\omega_{j}B_{0,\alpha}^{j}(t)-i\,g_{0,\alpha}^{*}\exp\left(-i\:\omega_{0,\alpha}t\right)\\
 &  & -\int_{0}^{t}\text{d}\tau\,f_{j}(t-\tau)B_{0,\alpha}^{j}(\tau)-\int_{0}^{t}\text{d}\tau\,f_{0}(t-\tau)\sum_{p=1,2}B_{0,\alpha}^{p}(\tau),\\
\dot{B}_{1,\alpha}^{j}(t) & = & -i\,\omega_{j}B_{1,\alpha}^{j}(t)-i\,g_{1,\alpha}^{*}\exp\left(-i\:\omega_{1,\alpha}t\right)\delta_{j1}\\
 &  & -\int_{0}^{t}\text{d}\tau\,f_{j}(t-\tau)B_{1,\alpha}^{j}(\tau)-\int_{0}^{t}\text{d}\tau\,f_{0}(t-\tau)\sum_{p=1,2}B_{1,\alpha}^{p}(\tau),\\
\dot{B}_{2,\alpha}^{j}(t) & = & -i\,\omega_{j}B_{2,\alpha}^{j}(t)-i\,g_{2,\alpha}^{*}\exp\left(-i\:\omega_{2,\alpha}t\right)\delta_{j2}\\
 &  & -\int_{0}^{t}\text{d}\tau\,f_{j}(t-\tau)B_{2,\alpha}^{j}(\tau)-\int_{0}^{t}\text{d}\tau\,f_{0}(t-\tau)\sum_{p=1,2}B_{2,\alpha}^{p}(\tau).
\end{eqnarray*}
 The expansion coefficients can be calculated using the Laplace transform
method.  Applying the Laplace transform to the above equations, we
obtain
\begin{eqnarray*}
s\bar{C}_{1}^{1}-1 & = & -i\,\omega_{1}\bar{C}_{1}^{1}-\bar{f}_{1}\bar{C}_{1}^{1}-\bar{f}_{0}\left(\bar{C}_{1}^{1}+\bar{C}_{1}^{2}\right)\\
s\bar{C}_{1}^{2} & = & -i\,\omega_{2}\bar{C}_{1}^{2}-\bar{f}_{2}\bar{C}_{1}^{2}-\bar{f}_{0}\left(\bar{C}_{1}^{1}+\bar{C}_{1}^{2}\right)
\end{eqnarray*}
\begin{eqnarray*}
s\bar{C}_{2}^{1} & = & -i\,\omega_{1}\bar{C}_{2}^{1}-\bar{f}_{1}\bar{C}_{2}^{1}-\bar{f}_{0}\left(\bar{C}_{2}^{1}+\bar{C}_{2}^{2}\right)\\
s\bar{C}_{2}^{2}-1 & = & -i\,\omega_{2}\bar{C}_{2}^{2}-\bar{f}_{2}\bar{C}_{2}^{2}-\bar{f}_{0}\left(\bar{C}_{2}^{1}+\bar{C}_{2}^{2}\right)
\end{eqnarray*}
\begin{eqnarray*}
s\bar{B}_{0,\alpha}^{1} & = & -i\,\omega_{1}\bar{B}_{1,\alpha}^{0}-\frac{i\,g_{0,\alpha}^{*}}{s+i\,\omega_{0,\alpha}}-\bar{f}_{1}\bar{B}_{0,\alpha}^{1}-\bar{f}_{0}\left(\bar{B}_{0,\alpha}^{1}+\bar{B}_{0,\alpha}^{2}\right)\\
s\bar{B}_{0,\alpha}^{2} & = & -i\,\omega_{2}\bar{B}_{0,\alpha}^{2}-\frac{i\,g_{0,\alpha}^{*}}{s+i\,\omega_{0,\alpha}}-\bar{f}_{2}\bar{B}_{0,\alpha}^{2}-\bar{f}_{0}\left(\bar{B}_{0,\alpha}^{1}+\bar{B}_{0,\alpha}^{2}\right)
\end{eqnarray*}
\begin{eqnarray*}
s\bar{B}_{1,\alpha}^{1} & = & -i\,\omega_{1}\bar{B}_{1,\alpha}^{1}-\frac{i\,g_{1,\alpha}^{*}}{s+i\,\omega_{1,\alpha}}-\bar{f}_{1}\bar{B}_{1,\alpha}^{1}-\bar{f}_{0}\left(\bar{B}_{1,\alpha}^{1}+\bar{B}_{1,\alpha}^{2}\right)\\
s\bar{B}_{1,\alpha}^{2} & = & -i\,\omega_{2}\bar{B}_{1,\alpha}^{2}-\bar{f}_{2}\bar{B}_{1,\alpha}^{2}-\bar{f}_{0}\left(\bar{B}_{1,\alpha}^{1}+\bar{B}_{1,\alpha}^{2}\right)
\end{eqnarray*}
\begin{eqnarray*}
s\bar{B}_{2,\alpha}^{1} & = & -i\,\omega_{1}\bar{B}_{2,\alpha}^{1}-\bar{f}_{1}\bar{B}_{2,\alpha}^{1}-\bar{f}_{0}\left(\bar{B}_{2,\alpha}^{1}+\bar{B}_{2,\alpha}^{2}\right)\\
s\bar{B}_{2,\alpha}^{2} & = & -i\,\omega_{2}\bar{B}_{2,\alpha}^{2}-\frac{i\,g_{2,\alpha}^{*}}{s+i\,\omega_{2,\alpha}}-\bar{f}_{2}\bar{B}_{2,\alpha}^{2}-\bar{f}_{0}\left(\bar{B}_{2,\alpha}^{1}+\bar{B}_{2,\alpha}^{2}\right)
\end{eqnarray*}
which leads to solutions
\begin{eqnarray*}
\bar{C}_{1}^{1} & = & \frac{\bar{f}+s+i\,\omega_{2}}{\left(\bar{f}+s\right)\,\left(\bar{f}+s+i\,\left(\omega_{1}+\omega_{2}\right)\right)-\omega_{1}\,\omega_{2}-\bar{f}^{2}\,\cos^{4}\theta},\\
\bar{C}_{1}^{2} & = & -\frac{\bar{f}\,\cos^{2}\theta}{\left(\bar{f}+s\right)\,\left(\bar{f}+s+i\,\left(\omega_{1}+\omega_{2}\right)\right)-\omega_{1}\,\omega_{2}-\bar{f}^{2}\,\cos^{4}\theta},
\end{eqnarray*}
\begin{eqnarray*}
\bar{C}_{2}^{1} & = & -\frac{\bar{f}\,\cos^{2}\theta}{\left(\bar{f}+s\right)\,\left(\bar{f}+s+i\,\left(\omega_{1}+\omega_{2}\right)\right)-\omega_{1}\,\omega_{2}-\bar{f}^{2}\,\cos^{4}\theta},\\
\bar{C}_{2}^{2} & = & \frac{\bar{f}+s+i\,\omega_{1}}{\left(\bar{f}+s\right)\,\left(\bar{f}+s+i\,\left(\omega_{1}+\omega_{2}\right)\right)-\omega_{1}\,\omega_{2}-\bar{f}^{2}\,\cos^{4}\theta},
\end{eqnarray*}
\begin{eqnarray*}
\bar{B}_{0,\alpha}^{1} & = & -\frac{i\,g_{0,\alpha}^{*}\,\left(\bar{f}\,\sin^{2}\theta+s+i\,\omega_{2}\right)}{\left(s+i\,\omega_{0,\alpha}\right)\,\left(\left(\bar{f}+s\right)\,\left(\bar{f}+s+i\,\left(\omega_{1}+\omega_{2}\right)\right)-\omega_{1}\,\omega_{2}-\bar{f}^{2}\,\cos^{4}\theta\right)},\\
\bar{B}_{0,\alpha}^{2} & = & -\frac{i\,g_{0,\alpha}^{*}\,\left(\bar{f}\,\sin^{2}\theta+s+i\,\omega_{1}\right)}{\left(s+i\,\omega_{0,\alpha}\right)\,\left(2\left(\bar{f}+s\right)\,\left(\bar{f}+s+i\,\left(\omega_{1}+\omega_{2}\right)\right)-\omega_{1}\,\omega_{2}-\bar{f}^{2}\,\cos^{4}\theta\right)},
\end{eqnarray*}
\begin{eqnarray*}
\bar{B}_{1,\alpha}^{1} & = & -\frac{i\,g_{1,\alpha}^{*}\,\left(\bar{f}+s+i\,\omega_{2}\right)}{\left(s+i\,\omega_{1,\alpha}\right)\,\left(\left(\bar{f}+s\right)\,\left(\bar{f}+s+i\,\left(\omega_{1}+\omega_{2}\right)\right)-\omega_{1}\,\omega_{2}-\bar{f}^{2}\,\cos^{4}\theta\right)},\\
\bar{B}_{1,\alpha}^{2} & = & \frac{i\,g_{1,\alpha}^{*}\,\bar{f}\,\cos^{2}\theta}{\left(s+i\,\omega_{1,\alpha}\right)\,\left(\left(\bar{f}+s\right)\,\left(\bar{f}+s+i\,\left(\omega_{1}+\omega_{2}\right)\right)-\omega_{1}\,\omega_{2}-\bar{f}^{2}\,\cos^{4}\theta\right)},.
\end{eqnarray*}
\begin{eqnarray*}
\bar{B}_{2,\alpha}^{1} & = & \frac{i\,g_{2,\alpha}^{*}\,\bar{f}\,\cos^{2}\theta}{\left(s+i\,\omega_{2,\alpha}\right)\,\left(\left(\bar{f}+s\right)\,\left(\bar{f}+s+i\,\left(\omega_{1}+\omega_{2}\right)\right)-\omega_{1}\,\omega_{2}-\bar{f}^{2}\,\cos^{4}\theta\right)},\\
\bar{B}_{2,\alpha}^{2} & = & -\frac{i\,g_{2,\alpha}^{*}\,\left(\bar{f}+s+i\,\omega_{1}\right)}{\left(s+i\,\omega_{2,\alpha}\right)\,\left(\left(\bar{f}+s\right)\,\left(\bar{f}+s+i\,\left(\omega_{1}+\omega_{2}\right)\right)-\omega_{1}\,\omega_{2}-\bar{f}^{2}\,\cos^{4}\theta\right)}.
\end{eqnarray*}
Here, we have assumed that the two independent reservoirs have the
same spectral structure, with $f_{1}=f_{2}$, and that the relation
between the common reservoir and the independent reservoirs is given
by $\ensuremath{f_{0}=f\cos^{2}\theta}$and$f_{1}=f\sin^{2}\theta$.

\end{widetext}


\begin{thebibliography}{99}

\bibitem{Verstraete2009}
F.~Verstraete, M.~M.~Wolf, and J.~I.~Cirac,
Quantum computation and quantum-state engineering driven by dissipation,
Nature Physics \textbf{5}, 633--636 (2009).

\bibitem{Yosifov2024}
A.~Yosifov, A.~Iyer, D.~Ebler, and V.~Vedral,
Quantum homogenization as a quantum steady-state protocol on noisy intermediate-scale quantum hardware,
Phys. Rev. A \textbf{109}, 032624 (2024).

\bibitem{Wang2018}
Z.~Wang, W.~Wu, G.~Cui, and J.~Wang,
Coherence enhanced quantum metrology in a nonequilibrium optical molecule,
New J. Phys. \textbf{20}, 033034 (2018).

\bibitem{Naghiloo2019}
M.~Naghiloo, M.~Abbasi, Y.~N.~Joglekar, and K.~W.~Murch,
Quantum state tomography across the exceptional point in a single dissipative qubit,
Nature Physics \textbf{15}, 1232--1236 (2019).

\bibitem{Goold2016}
J.~Goold, M.~Huber, A.~Riera, L.~del~Rio, and P.~Skrzypczyk,
The role of quantum information in thermodynamics---a topical review,
J. Phys. A: Math. Theor. \textbf{49}, 143001 (2016).

\bibitem{Lu2024}
J.~Lu, Z.~Wang, J.~Ren, C.~Wang, and J.-H.~Jiang,
Electron-phonon nonequilibrium thermodynamics beyond local equilibrium: Effective chemical potential of phonons and giant Seebeck coefficient,
Phys. Rev. B \textbf{109}, 125407 (2024).

\bibitem{Ask2022}
A.~Ask and G.~Johansson,
Non-Markovian steady states of a driven two-level system,
Phys. Rev. Lett. \textbf{128}, 083603 (2022).

\bibitem{Ablimit2024}
A.~Ablimit, Z.~M.~Wang, F.~H.~Ren, P.~Brumer, and L.~A.~Wu,
Driven open double two-level systems: Exact non-Markovian dynamics and entanglement generation,
Phys. Rev. A \textbf{110}, 052220 (2024).

\bibitem{Singh2021}
D.~Singh,
Origin of loose bound of the thermodynamic uncertainty relation in a dissipative two-level quantum system,
Phys. Rev. E \textbf{104}, 054115 (2021).

\bibitem{Becker2022}
T.~Becker, A.~Schnell, and J.~Thingna,
Thermodynamics with local master equations,
Phys. Rev. Lett. \textbf{129}, 200403 (2022).

\bibitem{Tucker2020}
K.~Tucker, D.~Barberena, R.~J.~Lewis-Swan, J.~K.~Thompson, J.~G.~Restrepo, and A.~M.~Rey,
Facilitating spin squeezing generated by collective dynamics with single-particle decoherence,
Phys. Rev. A \textbf{102}, 051701 (2020).

\bibitem{Liu2024}
Y.~Liu, W.-B.~Yan, Y.~J.~Xia, and Z.-X.~Man,
The effects of common reservoirs on the performance of a quantum refrigerator,
J. Phys. A: Math. Theor. \textbf{57}, 285301 (2024).

\bibitem{Venkatesh2018}
B.~P.~Venkatesh, M.~L.~Juan, and O.~Romero-Isart,
Quantum-limited measurement of a mechanical oscillator,
Phys. Rev. Lett. \textbf{120}, 033602 (2018).

\bibitem{Guarnieri2018}
G.~Guarnieri, M.~Kol\'{a}\v{r}, and R.~Filip,
Hybrid-environment quantum metrology in open quantum systems,
Phys. Rev. Lett. \textbf{121}, 070401 (2018).

\bibitem{Gribben2020}
D.~Gribben, A.~Strathearn, J.~Iles-Smith, D.~Kilda, A.~Nazir, B.~W.~Lovett, and P.~Kirton,
Exact quantum dynamics in structured environments,
Phys. Rev. Res. \textbf{2}, 013265 (2020).

\bibitem{Zhou2022}
Y.~Zhou, J.~Hu, and H.~Yu,
Conditions for steady-state entanglement of quantum systems in a stationary environment under Markovian dissipation,
Phys. Rev. A \textbf{105}, 032426 (2022).

\bibitem{Zanardi2014}
P.~Zanardi and L.~Campos~Venuti,
Coherent quantum dynamics in steady-state manifolds of strongly dissipative systems,
Phys. Rev. Lett. \textbf{113}, 240406 (2014).

\bibitem{Dodin2021}
A.~Dodin and P.~Brumer,
Noise-induced coherence in molecular processes,
J. Phys. B: At. Mol. Opt. Phys. \textbf{54}, 223001 (2021).

\bibitem{Dodin2024}
A.~Dodin, T.~V.~Tscherbul, and P.~Brumer,
Non-Hermitian quantum effects in molecular spectroscopy and dynamics,
J. Phys. Chem. Lett. \textbf{15}, 7694--7705 (2024).

\bibitem{Wu2002}
L.-A.~Wu and D.~A.~Lidar,
Creating decoherence-free subspaces using strong and fast pulses,
Phys. Rev. Lett. \textbf{88}, 207902 (2002).

\bibitem{Pushin2011}
D.~A.~Pushin, M.~G.~Huber, M.~Arif, and D.~G.~Cory,
Experimental realization of decoherence-free subspace in neutron interferometry,
Phys. Rev. Lett. \textbf{107}, 150401 (2011).

\bibitem{Mewes2005}
C.~Mewes and M.~Fleischhauer,
Decoherence in collective quantum memories for photons,
Phys. Rev. A \textbf{72}, 022327 (2005).

\bibitem{Ruimy2021}
R.~Ruimy, A.~Gorlach, C.~Mechel, N.~Rivera, and I.~Kaminer,
Toward atomic-resolution quantum measurements with coherently shaped free electrons,
Phys. Rev. Lett. \textbf{126}, 233403 (2021).

\bibitem{Eastham2016}
P.~R.~Eastham, P.~Kirton, H.~M.~Cammack, B.~W.~Lovett, and J.~Keeling,
Bath-induced coherence and the secular approximation,
Phys. Rev. A \textbf{94}, 012110 (2016).

\bibitem{PhysRevA.32.2462}
F.~Haake and R.~Reibold,
Strong damping and low-temperature anomalies for the harmonic oscillator,
Phys. Rev. A \textbf{32}, 2462 (1985).

\bibitem{PhysRevD.45.2843}
B.~L.~Hu, J.~P.~Paz, and Y.~Zhang,
Quantum Brownian motion in a general environment: Exact master equation,
Phys. Rev. D \textbf{45}, 2843 (1992).

\bibitem{PhysRevE.55.153}
R.~Karrlein and H.~Grabert,
Exact time evolution and the driven damped quantum oscillator,
Phys. Rev. E \textbf{55}, 153--164 (1997).

\bibitem{PhysRevLett.109.170402}
W.-M.~Zhang, P.-Y.~Lo, H.-N.~Xiong, M.~W.-Y.~Tu, and F.~Nori,
General non-Markovian dynamics of open quantum systems without rotating-wave approximation,
Phys. Rev. Lett. \textbf{109}, 170402 (2012).

\bibitem{arXiv:2211.15722}
M.~Lampo and M.~B.~Plenio,
Quantum Brownian motion with spatially correlated noise,
arXiv:2211.15722 (2022).

\bibitem{PhysRevA.111.062206}
Z.~M.~Wang, S.~L.~Wu, M.~S.~Byrd, and L.~A.~Wu,
Analytical solutions of bilinear bosonic models and exact characterization of non-Markovian dynamics,
Phys. Rev. A \textbf{111}, 062206 (2025).

\bibitem{Wang2023}
Z.~M.~Wang, F.~H.~Ren, M.~S.~Byrd, and L.~A.~Wu,
Effect of quantum jumps on non-Hermitian systems,
Phys. Rev. A \textbf{108}, 022607 (2023).

\bibitem{Braun2011}
D.~Braun and J.~Martin,
Decoherence control by quantum bath engineering,
Nature Communications \textbf{2}, 223 (2011).

\bibitem{Molinares2025}
H.~Molinares and V.~Eremeev,
Drive-Loss Engineering and Quantum Discord Probing of Synchronized Optomechanical Squeezing,
Mathematics \textbf{13}(13), 2171 (2025).

\bibitem{Chou2008}
C.~H.~Chou, T.~Yu, and B.~L.~Hu,
Exact master equation and quantum decoherence of two coupled harmonic oscillators,
Phys. Rev. E \textbf{77}, 011112 (2008).

\bibitem{Strasberg2016}
P.~Strasberg, G.~Schaller, N.~Lambert, and T.~Brandes,
Nonequilibrium thermodynamics in the strong coupling and non-Markovian regime based on a reaction coordinate mapping,
New J. Phys. \textbf{18}, 073007 (2016).

\bibitem{Hartmann2020}
R.~Hartmann and W.~T.~Strunz,
Accuracy assessment of perturbative master equations: Embracing nonpositivity,
Phys. Rev. A \textbf{101}, 012103 (2020).

\bibitem{Weiss2021}
U.~Weiss,
Quantum Dissipative Systems, 4th ed.,
World Scientific, Singapore (2021).

\bibitem{RevModPhys.92.041002}
M.~Esposito, M.~A.~Ochoa, and M.~Galperin,
Quantum thermodynamics: A nonequilibrium Green's function approach,
Rev. Mod. Phys. \textbf{92}, 041002 (2020).

\bibitem{Onsager1933}
L.~Onsager,
Theories of concentrated electrolytes,
Chem. Rev. \textbf{13}, 73--89 (1933).

\bibitem{Campisi2009}
M.~Campisi, P.~Talkner, and P.~H\"anggi,
Fluctuation theorem for arbitrary open quantum systems,
Phys. Rev. Lett. \textbf{102}, 210401 (2009).

\bibitem{Subasi2012}
S.~Hilt, B.~Thomas, and E.~Lutz,
Hamiltonian of mean force for damped quantum systems,
Phys. Rev. E \textbf{84}, 031110 (2011).

\bibitem{Hilt2011}
S.~Hilt, B.~Thomas, and E.~Lutz,
Hamiltonian of mean force for damped quantum systems,
Phys. Rev. E \textbf{84}, 031110 (2011).

\bibitem{Brown2022}
M.~E.~Kimchi\textendash Schwartz, L.~Martin, E.~Flurin, C.~Aron, M.~Kulkarni, H.~E.~T{\"u}reci, and I.~Siddiqi,
Stabilizing Entanglement via Symmetry\textendash Selective Bath Engineering in Superconducting Qubits,
Phys. Rev. Lett. \textbf{116}, 240503 (2016).

\bibitem{Hatridge2024}
M.~Hatridge, S.~Shankar, M.~Mirrahimi, K.~W.~Murch, R.~J.~Schoelkopf, and M.~H.~Devoret,
Hardware-efficient stabilization of entanglement via engineered dissipation in superconducting circuits,
Phys. Rev. Research \textbf{7}, L022018 (2025).

\end{thebibliography}
\end{document}